\documentclass[useAMS,usenatbib]{mn2e}

\usepackage{placeins}

\usepackage{graphicx} 
\usepackage{multirow}
\usepackage{array}

\usepackage{amsmath,amssymb}

\usepackage[T1]{fontenc}
\usepackage[latin9]{inputenc}
\usepackage{float}
\usepackage{graphicx}
\usepackage{esint}

\makeatletter

\DeclareFontEncoding{LGR}{}{}
\DeclareTextSymbol{\~}{LGR}{126}

\begin{document}

\title {A one-parameter formula for testing slow-roll dark energy: observational prospects} 

\author[Zachary Slepian, J. Richard Gott, III, and Joel Zinn]{Zachary Slepian\thanks{E-mail: zslepian@cfa.harvard.edu}, J. Richard Gott, III\thanks{E-mail: jrg@astro.princeton.edu}, and Joel Zinn\thanks{E-mail: jzinn@princeton.edu}\\ 
Harvard-Smithsonian Center for Astrophysics, 60 Garden St., Cambridge, MA 02138 (ZS)\\
Department of Astrophysical Sciences, Princeton University, Princeton, NJ 08544 (JRG, JZ)\\
}
\maketitle

\begin{abstract}
Numerous upcoming observations, such as WFIRST, BOSS, BigBOSS, LSST, $Euclid$, and $Planck$, will constrain dark energy (DE)'s equation of state with great precision.  They may well find the ratio of pressure to energy density, $w$, is $-1$, meaning DE is equivalent to a cosmological constant. However, {\bf many time-varying DE models have also been proposed.  A single parametrization to test a broad class of them and that is itself motivated by a physical picture is therefore desirable.}  We suggest the simplest model of DE has the same mechanism as inflation, likely a scalar field slowly rolling down its potential. If this is so, DE will have a generic equation of state and the Universe will have a generic  dependence of the Hubble constant on redshift independent of the potential's starting value and shape. {\bf This equation of state and expression for the Hubble constant offer the desired model-independent but physically motivated parametrization, because they will hold for most of the standard scalar-field models of DE such as quintessence and phantom DE}.  Up until now two-parameter descriptions of $w$ have been available, but this work finds an additional approximation that leads to a single parameter model. Using it, we conduct a $\chi^2$ analysis and find that {\bf experiments in the next seven years should be able to distinguish any of these time-varying DE models on the one hand from a cosmological constant on the other} to $73\%$ confidence if $w$ today differs from $-1$ by $3.5\%$. In the limit of perfectly accurate measurements of $\Omega_m$ and $H_0$, this confidence  would rise to $96\%$. We also include discussion of the current status of DE experiment, a table compiling the techniques each will use,  and tables of the precisions of the experiments for which this information was available at the time of publication.
\end{abstract}


\section{Introduction}
The Universe is now undergoing an accelerated expansion of space itself, driven by  dark energy (DE), a negative pressure component  that constitutes $73\%$ of the current energy density (Komatsu et al. 2011, Riess et al. 1998, Perlmutter et al. 1999, 1998).  While there is much evidence for DE's existence, we at present lack insight into its essence.  Observational efforts have focused on measuring its equation of state, parametrized by $w\equiv p/\rho=$ pressure/energy density, in the hope that so doing will offer understanding of DE's essential nature.

\newpage

Faced with a proliferation of theoretical models, observers have used parametrizations, such as that of Chevallier-Polarski-Linder, $w=w_0+w_a(1-a)$, where $a$ is the scale factor, to assess proposed programmes (e.g. Albrecht et al. 2006; Chevallier \& Polarski 2001, Linder 2003; for model compendia, see Copeland et al. 2006 and Li et al. 2011).  These parametrizations are supposed to allow a wide variety of models to be tested, since models can produce predictions for e.g. $w_0$ and $w_a$.  However, these parametrizations also build in a shape for $w(a)$, {\it one not based on any physical model.}  Therefore, it may be dangerous to design experiments and analyze results to look for these shapes, which there is no particular reason to believe $w(a)$ for DE in fact has.  But what is the alternative?  While there is no shortage of models, there is one of funding, time, and resources.

Ockham's razor would suggest that, all else equal, the simplest model is the best.  The simplest model, in our view, would take advantage of the fact that we have already most likely seen one epoch of accelerated expansion: inflation in the early universe (Guth 1981, Guth \& Kaiser 2005, Guth 2007).   The idea that DE now is produced by the same mechanism as inflation then is simple---and it immediately suggests certain dark energy models are implausible.\footnote{For instance, we would argue that dark energy is unlikely to be driven by a phantom field. Inflation certainly was not driven by a phantom field, for had it been we would not be here, as a `big rip' would have occurred already (Caldwell 2002, Caldwell et al. 2003).  Importantly, this comment also applies to the current `standard' model of dark energy, as a constant vacuum energy density (cosmological constant) with $w\equiv -1$. In this model, the current Universe sits stationary at the bottom of a potential well with a value of $V_0=10^{-120}$ in Planck units (Weinberg 1987, Vilenkin 2003). However, if DE $is$ just another epoch of inflation, neither inflation nor DE can be due to a constant energy density, or we would not be here---the Universe would still be undergoing high-speed inflation!}

There are a variety of models for inflation, but a popular, simple class of models that seems to fit all observational constraints has inflation being driven by a scalar field in slow-roll (Linde 1982, Albrecht \& Steinhardt 1982).  Indeed, values of the scalar spectral tilt $n_s$ less than unity are a hallmark of slow-roll inflation that is in fact observed (Linde 2002, Komatsu et al. 2011).  In a previous paper (Gott \& Slepian 2011), we focused on a specific slow-roll inflationary (and, by analogy, DE) model: the quadratic potential associated with Linde's chaotic inflation.  This model for inflation is observationally allowed, and has the potential to be confirmed in the near future if Planck detects the tensor mode amplitude $r$ the model predicts (for Planck details, see Planck: the scientific programme 2005; Geisbuesch \& Hobson 2006; de Putter et al. 2009).  

Nonetheless, at present, while slow-roll itself is a compelling and broadly accepted model for inflation, the details of {\bf the shape of the potential} and {\bf starting value of the scalar field} are unknown; this is also the case for phantom DE, to which we show our results apply.  Consequently, in this paper, we simplify and generalize our previous work (Gott \& Slepian 2011) by exploring the observational signature of DE as a scalar field slowly rolling down its potential.\footnote{For review of previous dark energy models, with a focus on those most similar to that proposed here, see Gott \& Slepian 2011.  A brief review of the evidence for inflation may also be found there.}  We find the generic result that if DE is a scalar field in slow-roll, $w+1$ evolves with redshift proportionally to $1/H^2$, with $w$ the ratio of pressure to energy density in the DE and $H$ the Hubble constant (also shown in Gott \& Slepian 2011).  This result is independent of the initial value and shape of the potential as long as the slow-roll conditions are met (see \S2.1).  We apply this scaling to find a closed-form approximate formula for the Hubble constant in a slow-roll DE cosmology. This makes it easy to find the luminosity and angular diameter distances and chart their differences from those for a $w \equiv -1$ cosmology. These differences have a characteristic shape: the observational signature of slow-roll dark energy.

The paper is laid out as follows. In the remainder of this Introduction (\ref{subsec:prevwork}), we offer a brief overview of pervious work on parametrizing and constraining scalar field models of DE similar to those discussed here. In \S\ref{sec:closedform}, we derive formulae for the DE equation of state and the Hubble constant if DE is a scalar field slowly rolling down its potential, and show that, in addition to quintessence, these formulae apply to phantom DE.  In \S\ref{sec:observing}, we review common techniques for observing DE, and in \S\ref{subsec:currentandupcoming} turn to current and future experiments.  \S5 outlines our method for determining the confidence with which slow-roll DE (be it quintessence or phantom) may be distinguished from a cosmological constant, carrying out a $\chi^2$ analysis first neglecting errors in the cosmological parameters and then accounting for them. \S6 presents and discusses the results of this latter calculation, while \S\ref{sec:conclusion} concludes by placing our work in context and recapitulating the paper's central points.

An Appendix (\S10) presents discussion of the self-consistency and accuracy of the approximations of \S2 and some illustrative numerical results from exact solution of the field equation of motion and the Friedmann equation for typical potentials in the DE models we consider.  We also provide low-redshift expressions for the fractional differences in Hubble constant and comoving distance between slow-roll and cosmological constant cosmologies, scalings for the $\chi^2$'s we will have earlier (\S5) computed numerically, and finally, the method we will have used to reconstruct the coefficients of the confidence ellipse presented in \S6. A second Appendix (\S11) comprises tables of the experimental precisions of WFIRST, BigBOSS, Euclid, LSST, and BOSS as used in \S5.

The central points of the paper are as follows:
\begin{enumerate}
\item If  DE is not equivalent to a cosmological constant, then we suggest it is likely a scalar field in slow-roll by analogy with inflation.  Even if it is phantom DE:
\item It will have a generic equation of state given by eqn. (9) and a generic $H(z)$ given by eqn. (14).
\item  With this generic $H(z)$, observations in the next seven years can in principle distinguish between these forms of time-varying DE on the one hand and a cosmological constant on the other.\footnote{Our results will hold for any field where DE is provided by the scalar field's potential and the slow-roll conditions are met; importantly, our results will not apply to k-essence models, where DE is provided by the scalar field's kinetic energy (cf. Armendariz-Picon et al. 2001). For discussion of slow-roll thawing k-essence models see Chiba et al. 2009.}
\item  Neglecting errors in the cosmological parameters and taking the difference from $w=-1$ at present to be $3.5\%$, this can be done to $96\%$ confidence. The confidence levels of a possible detection neglecting errors for other values of $w$ today are illustrated in Figure 4.
\item For $w+1 = 3.5\%$ today and accounting for the possibility of errors in the cosmological parameters, slow-roll DE may be distinguished from cosmological constant DE with $73\%$ confidence.  The confidence levels of a possible detection for other values of $w$ today are illustrated in Figure 11.
\end{enumerate}

\subsection{Previous parametrizations \& constraints for slow-roll DE}
\label{subsec:prevwork}
Numerous authors have worked to develop parametrizations of slow-roll DE, as well as used observations to place constraints on these models.  Here we provide a brief review of work closely bearing on our own; for a more extensive treatment see Chiba et al. 2013, Chiba et al. 2009, and Gott \& Slepian  2011.  Dutta \& Scherrer (2008) provided an expansion for a scalar field with $w\approx -1$ rolling near a local maximum in its potential.  This extended earlier work by Scherrer \& Sen (2008) by generalizing to potentials with non-zero curvature.  Crittenden et al. (2007), Neupane \& Scherer (2008), and Cahn et al. (2008) offered other early attempts to develop both appropriate slow-roll conditions and parametrizations for $w$.  Chiba (2009) represents a continuation of these efforts, deriving general slow-roll conditions which allow a two-parameter form for $w$ that further generalizes the results of Dutta \& Scherrer 2008.\footnote{Chiba et al. 2010 obtain a similar parametrization for a non-minimally coupled scalar field.}   Both Chiba (2009) and Dutta \& Scherrer (2008) make the point that slow-roll DE models are in general poorly described by the Chevallier-Polarski-Linder parametrization $w(z)=w_0+w_a(1-a)$, a comment we emphasize again here.

Overall, then, on the theory side, previous work had succeeded in reducing slow-roll quintessence models to an equation of state $w$ described by two parameters.  The main advance of this work is to further collapse this representation to a one-parameter class of models.  Previous work required an additional parameter because it effectively described the value of the acceleration in the equation of motion, or equivalently, the curvature of the potential.  This was deemed necessary because, in contrast with the inflationary case, the acceleration cannot always be taken to be small.  However, in this work we show that as long as the acceleration is either 1) small or 2) roughly constant in time, it will not affect the scaling of $w$ with time, meaning it can be eliminated from the problem.  This argument is made in a physically intuitive way in \S2, and we take a more mathematically rigorous approach, as well as quantifying the error introduced by this approximation in \S10.  Figure 19 in particular shows that the error our approximation introduces should be at least an order of magnitude lower than the signal we expect to use to distinguish slow-roll DE from a cosmological constant for the representative case of $V\propto \phi^2$. We also obtain more general  analytic expressions for this error which suggest that this should hold for other potentials.

We now turn to a brief review of prior work placing observational constraints on parametrizations of slow-roll DE.   The most comprehensive work to date is Chiba et al. (2013), which puts bounds on the parameters for $w$ using supernovae type Ia (SNIa), the Cosmic Microwave Background (CMB), and the Baryon Acoustic Oscillation (BAO).  For thawing models (models where the field has begun to involve only at late times; see Caldwell \& Linder 2005) they use Chiba's (2009) two-parameter model, while for tracking freezing models (where the field has now frozen to a halt and different initial conditions converge to a common trajectory (tracker) (Chiba et al. 2013)) they use the two parameter form of Chiba 2010.  For  scaling freezing models (the equation of state scales with the background fluid's) numerical simulations are employed.  Previously, Dutta \& Scherrer (2008) and Chiba et al. (2009) had done this analysis for thawing models with smaller datasets, while Chiba (2010)  and Wang et al. (2012) did similarly for tracking feezing models. Novosyadlyj et al. (2011) do such an analysis using a two-parameter model developed in Novosyadlyj et al. 2010. Since our formula for $w$ has one rather than two parameters, these results unfortunately do not give much insight into the most likely values of $\delta w_0=w+1$ if $w$ follows the formula we obtain. In future we hope to use the data sets available to do a similar analysis to constrain this formula.

\section{Slow-roll scalar field DE}
\label{sec:closedform}

\subsection{Equation of state}
\label{subsec:eos}
We begin by defining $\delta w=w+1$: simply the difference between $w$ and negative one, a worthwhile subject of attention because it would be interesting if observation finds $w\neq -1$ (or $\delta w\neq0$).  We currently know that $w\approx -1$ to an accuracy of approximately $7\%$ (Komatsu et al. 2011).  If future observations tighten the limits around $w = -1$, that will increase confidence in the standard cosmological constant model, but will not tell us anything new about DE.  

The most exciting result of future observations would be to find a value of $w\neq -1$, for then we would learn something new about dark energy.  Of all the models of dark energy with $w\neq -1$, we would argue that slow-roll DE is the most conservative, since we have seen that behavior in the Universe before during inflation.  

If one hopes to detect a significant deviation of $w$ from $-1$, it is helpful to know the functional form $w(z)$ is likely to take. That will be our goal here.  

In what follows, when discussing the scalar field and its potential, we will work in Planck units, $c=\hbar =8\pi G =1$. 

 For a scalar field $\phi$ with potential $V(\phi$), we have pressure $p=\frac{1}{2}\dot{\phi}^2-V(\phi)$ and energy density $\rho=\frac{1}{2}\dot{\phi}^2+V(\phi)$, where ``dot'' denotes a time derivative.  Hence
\begin{equation}
w\equiv \frac{p}{\rho}=\frac{\frac{1}{2}\dot{\phi}^2-V(\phi)}{\frac{1}{2}\dot{\phi}^2+V(\phi)}
\label{e:wscalfield}
\end{equation}
and so 

\begin{equation}
\delta w=\frac{\dot{\phi}^{2}}{\frac{1}{2}\dot{\phi}^{2}+V}.
\end{equation}

Evidently, small $\delta w$ at present implies $\dot{\phi}^2  \ll V$ at present.  To connect this with the traditional analysis of a slow-roll field in inflation, note that the condition $\dot{\phi}  \ll V$ (which is implied by the above) is just the first of the two slow-roll conditions usually imposed (see e.g. Lesgourgues 2006).  In inflationary cosmology, this condition comes from the need for a roughly constant energy density to drive exponential expansion.  Since this energy density is offered by the scalar field's potential, that potential has to be nearly constant, and so $\dot{\phi}$ must be small compared to $V(\phi)$. The same applies for DE, as we observe exponential expansion today.   Furthermore, we know $w$ is not very much different from $-1$ for DE now, so if DE is a scalar field, then to be observationally allowed, $\delta w$ must be small now.  Thus we must have $\dot{\phi}^2  \ll V$ now. 

With this in hand, we approximate that
\begin{equation}
\delta w\approx\frac{\dot{\phi}^{2}}{V}.
\label{e:three}
\end{equation}

The scalar field evolves according to the equation of motion
\begin{equation}
\ddot{\phi}+3H\dot{\phi}=-\frac{\partial V}{\partial \phi}.
\label{e:eom}
\end{equation}
(see e.g. Linde 2002 or Copeland et al. 2006).  

We may rewrite this as
\begin{equation}
3H\dot{\phi}=-\frac{\partial V}{\partial \phi}\left[1+SR2\right]\nonumber,
\label{e:eomrewrite}
\end{equation}
with $SR2\equiv\ddot\phi/\left(\partial V/\partial \phi \right)$.

Note the analogy of eqn. (\ref{e:eom}) to that for a ball rolling down a hill with a frictional force.  In such a situation the ball quickly reaches terminal velocity and $\ddot{\phi}$ becomes small.  With this in mind, it should be intuitively clear why in slow-roll $inflationary$ models  $\ddot{\phi}$ is small (see e.g. Copeland et al. 2006 or Lyth \& Liddle 2000).  This is traditionally the second of the two slow-roll conditions imposed in inflationary cosmology, and, in that context, is needed to guarantee that the first ($\dot{\phi}^2  \ll V$) holds in general and not just at one particular time (Lesgourgues 2006).

We, however, require a slightly $less$ restrictive condition to obtain a scaling for $\dot\phi$.\footnote{As we discuss in \S1.1 and \S7, there has been much previous work to derive the analogs of the inflationary slow-roll conditions for scalar field DE; particularly the work of Chiba (2009) giving the slow-roll DE analogs of the inflationary parameters $\epsilon\equiv 1/2(V'/V)^2\ll 1$ and $\eta\equiv V''/V \ll 1$ (equivalent to $SR1\equiv \dot{\phi}^2/2V\ll1$ and $|SR2|\equiv \big|\ddot{\phi}/(\partial V/\partial \phi)\big| \ll1$).  He shows that $\big|\beta\big| \equiv \big| \ddot{\phi}/3H\dot{\phi}\big|$ is either negligible compared to unity (for freezing models; see Caldwell \& Linder 2005) or roughly constant (for thawing models).  The EOM gives $\dot{\phi}=-V'/3(1+\beta)H$, so for either case as long as $V'$ is roughly constant we recover $\dot{\phi}\propto 1/H$, as we would expect since $SR2$ contains information about $\ddot{\phi}$, as does $\beta$.  In contrast to this work, Chiba $does$ $not$ argue that $V'$ is roughly constant, nor that this, in conjunction with the behavior of $\beta$, means $\dot{\phi}\propto 1/H$.}
  Notice that, as long as $SR2$ is either 1) small compared to unity or 2) roughly constant in time, we will have 

 \begin{equation}
3H\dot{\phi}\propto- \frac{\partial V}{\partial\phi}.
\label{e:approxeom}
\end{equation}

If the scaling were an equality, this would duplicate the standard result in inflationary cosmology (see e.g. Lesgourgues 2006).  

Now, as long as the velocity $\dot \phi$ is small compared to the potential $V$ now and in the past,  $V$ will be nearly constant in time.  We know that the velocity is small now because $\delta w\propto \dot\phi^2$ is small now. Recalling the ball rolling down a hill analogy, it is clear that $|\dot\phi|$ is a monotonically increasing function, so $|\dot\phi|$ was even smaller in the past. Therefore from eqn. (\ref{e:three}) 

\begin{equation}
\delta w\propto \dot{\phi}^2.
\end{equation}

Now, $\frac{\partial V}{\partial \phi}$ is just some function of $\phi$ (for a quadratic potential, it is linear in $\phi$, for instance), and since
as we have already noted, $\phi$ is approximately constant in time, $\frac{\partial V}{\partial \phi}$ is also approximately constant in time.  Hence $3H\dot{\phi}\approx constant$.  Thus, from eqn. (\ref{e:approxeom})

To sho\begin{equation}
\dot{\phi}\propto\frac{1}{H}.
\label{e:phidotscaling}
\end{equation}

Since $\delta w\propto\dot{\phi}^{2}$, eqn. (\ref{e:phidotscaling}) implies that
\begin{equation}
\delta w\propto\frac{1}{H^{2}}.
\end{equation}

Normalizing appropriately, we obtain

\begin{equation}
\delta w(z)\approx\delta w_{0}\left(\frac{H_{0}}{H(z)}\right)^{2},
\label{e:slowroll}
\end{equation}
where, as will be true throughout this work, subscript nought denotes a quantity's value today.  This result was obtained in Gott \& Slepian 2011, and verified as a good approximation in the case of a $V=\frac {1}{2} m^2\phi^2$ potential by self-consistent, exact numerical solution of the equation of motion and the Friedmann equation.  We present this result, and numerical results for several other potentials, in Appendix A, \S10.1.  These bear out that for typical values of $\delta w\lesssim 5\%$, the field will be in slow-roll and our formula $\delta w\propto H^{-2}$ will fit reasonably well. Further discussion of the self-consistency of the approximations leading to this formula also occurs in \S10.1.  A more detailed and mathematically rigourous treatment of the validity of the two slow-roll approximations we have used (the second, as we have noted, being a less stringent version of the condition typically employed in inflationary cosmology) follows in Appendix A, \S10.2. 

In summary, then, for small $\delta w$, slow-roll applies, $\dot{\phi}$
is small, and $\delta w\propto\dot{\phi}^{2}.$ When the field
reaches terminal velocity, the acceleration $\ddot{\phi}$ is
nearly zero, so the Hubble friction term $3H\dot{\phi}$ is balanced
by the slope of the potential the field is rolling down, $-\partial V/\partial\phi$.
This is analogous to a ball rolling down a hill: it will reach terminal
velocity when energy dissipation from friction cancels out the energy
it gains by moving to lower values of its potential. Finally, if $\dot{\phi}$ is small, $\phi$
is roughly constant in time, so $\partial V/\partial\phi$ will be roughly constant 
as well.  This means $\dot{\phi}\propto H^{-1}$, which leads to eqn. (\ref{e:slowroll}). This derivation
is valid for any smooth scalar field potential for $\delta w$ sufficiently small compared to unity.

Note finally that the field we have considered above is essentially a quintessence model of dark energy.  For a review of recent work on quintessence DE, see e.g. De Boni et al. 2011 and references therein or Novosyadlyj et al. 2010.

\subsection{Hubble constant}
\label{subsec:Hubbleconst}
We would like a closed form expression for $H$ in a slow-roll DE cosmology that is
explicitly in terms of $z$ alone. We obtain this below. Neglecting the energy density in radiation today, $\Omega_{r}\equiv \rho_r/\rho_{\rm crit}$, with $\rho_{\rm crit}\equiv 3H_0^2/8\pi G=3H_0^2$ in Planck units (where $8\pi G=1$ and $H_0$ is in units defined by the condition $t_{pl}=1$), for a flat cosmology the Friedmann equation is
\begin{eqnarray}
H^{2}(z)&=&H_{0}^{2}\bigg(\Omega_{m}(1+z)^{3}\nonumber\\
\nonumber\\
&+&\Omega_{DE}\exp\left[3\int_{0}^{z}\frac{dz'}{1+z'}\delta w(z')\right] \bigg{)}.
\label{e:Fmaneqnz}
\end{eqnarray}
Clearly, to obtain $H$ in closed form, 
we require $\delta w(z)$. We can use eqn. (\ref{e:slowroll}) as a first guess for $\delta w(z)$ in eqn. (\ref{e:Fmaneqnz}),
but what can we use for $H$ in eqn. (\ref{e:slowroll})? Since $\delta w  \ll 1,$ to a good
approximation $w=-1$, in which case we can write that\footnote{Note that if $\delta w\propto H^{-2}$, as we have shown, the approximation $\delta w \ll 1$ becomes even better in the past because $H$ rises in the past.}

\begin{equation}
H^{2}(z)\approx H_{0}^{2}\left(\Omega_{m}(1+z)^{3}+\Omega_{DE}\right).
\label{e:Happrx1}
\end{equation}

Using this for $H^2$ in eqn. (\ref{e:slowroll}), we have

\begin{equation}
\delta w(z)=\frac{\delta w_{0}}{\Omega_{m}(1+z)^{3}+\Omega_{DE}}, 
\label{e:deltawsubst}
\end{equation}

which we may now use in eqn. (\ref{e:Fmaneqnz}) to get an $H$ that incorporates the slow-roll nature of the DE.  Evaluating the integral in eqn. (\ref{e:Fmaneqnz}) that results from substituting using eqn. (\ref{e:deltawsubst}) for $\delta w$  leads to the first, approximate equality below, while computing this integral explicitly by partial fractions yields the second, exact equality.

\begin{eqnarray}
3\int_{0}^{z}\frac{dz'}{1+z'}\delta w(z')\approx3\int_{0}^{z}\frac{dz'}{1+z'}\left[\frac{\delta w_{0}}{\Omega_{m}(1+z')^{3}+\Omega_{DE}}\right] \nonumber \\
\nonumber \\
=\frac{\delta w_{0}}{\Omega_{DE}}\ln\left[\frac{(1+z)^{3}\left(\Omega_{m}+\Omega_{DE}\right)}{\Omega_{m}(1+z)^{3}+\Omega_{DE}}\right].
\label{e:lnforFeqn}
\end{eqnarray}

Substitution into eqn. (\ref{e:Fmaneqnz}) yields 

\begin{eqnarray}
H^{2}(z)&\approx& H_{0}^{2}\bigg(\Omega_{m}(1+z)^{3}\nonumber\\
\nonumber\\
&+&\Omega_{DE}\left[\frac{(1+z)^{3}}{\Omega_{m}(1+z)^{3}+\Omega_{DE}}\right]^{\delta w_{0}/\Omega_{DE}} \bigg),
\label{e:slowrollfman}
\end{eqnarray}
which is accurate to first order in $\delta w_0$.  We have used $\Omega_{m}+\Omega_{DE}\approx1$ to simplify the numerator of the argument of the logarithm in eqn. (\ref{e:lnforFeqn}) because we assume here and throughout this work a flat universe with negligible radiation density today.  This latter approximation means eqn. (\ref{e:slowrollfman}) is valid only for $z \ll 3196$, the redshift of matter-radiation equality.  Since observations are all done at much lower redshifts than $z=3196$, this is not a problematic restriction. 

Finally, we can substitute eqn. (\ref{e:slowrollfman}) into eqn. (\ref{e:slowroll}) to obtain a formula for $\delta w$ accurate to second order in $\delta w_0$.  So doing yields

\begin{eqnarray}
\delta w(z)&\approx&\delta w_0 \bigg(\Omega_{m}(1+z)^{3}\nonumber\\
&+&\Omega_{DE}\left[\frac{(1+z)^{3}}{\left(\Omega_{m}(1+z)^{3}+\Omega_{DE}\right)}\right]^{\delta w_{0}/\Omega_{DE}} \bigg) ^{-1}.
\label{e:moreaccslowroll}
\end{eqnarray}

\subsection{Application to phantom DE}
Phantom DE, also often referred to as "ghost" DE, has negative kinetic energy and leads to an equation of state $w\geq -1$ for the kinetic energy-dominated phase and $w\leq -1$ for the potential energy-dominated phase. These models reach a "big rip" singularity where the energy density becomes infinite in finite proper time; see Li et al. 2011 and Caldwell 2002 for further discussion, and Cline et al. 2004 for criticisms. For an example of a simple, minimal model of a phantom field, with non-canonical kinetic terms but no potential, see Chiba et al. (2000). For observational constraints on phantom DE models, see Novosyadlyj et al. 2012 and for forecasts of the possibility that future observations will distinguish between phantom DE and quintessence see Novosyadlyj et al. 2013.

For phantom DE, the equation of motion is 
\begin{equation}
\ddot{\phi}+3H\dot{\phi}=\frac{\partial V}{\partial \phi}.
\label{e:phantomeom}
\end{equation}
The difference from the scalar-field DE model as in eqn. (\ref{e:eom}) is that the right-hand side term here is positive: this means the field runs up the potential rather than rolling down it (Caldwell 2002).  Note that phantom DE has $p = -\frac{1}{2}\dot{\phi}^2-V(\phi)$ and $\rho =  -\frac{1}{2}\dot{\phi}^2 + V(\phi)$; i.e. phantom DE has a negative kinetic energy term. Following the same derivation as in \S2.1, it is easily shown that
\begin{equation}
\delta w_{\rm{ph}}\equiv w_{\rm{ph}}-(-1) \approx - \frac{\dot{\phi}^2}{V(\phi)},
\label{e:deltawphkey}
\end{equation}
where subscript ``$\rm{ph}$'' denotes "phantom".  The approximate equality is because we must have $\delta w_{\rm{ph}}  \ll 1$ today.  Comparing with eqn. (\ref{e:three}) shows that $\delta w_{0,\rm{ph}}\approx -\delta w_{0,\rm{q}}$, where subscript ``q'' denotes ``quintessence'', if we assume the same potential and magnitude of the velocity for each model.

Beginning with eqn. (\ref{e:deltawphkey}) and following the same derivation as in \S2.1 shows that
\begin{equation}
\delta w_{\rm{ph}}(z)\approx \delta w_{0,\rm{ph}}\left(\frac{H_0}{H(z)}\right)^2.
\label{e:deltawphz}
\end{equation}
Hence all of the formulae presented in this work for slow-roll scalar field DE will also be valid for phantom DE, but in this latter case,  $\delta w_0$ will be negative rather than positive.  For instance, most importantly, eqns. (14) and (16) apply to phantom DE, but both $\delta w_0$ and $\delta w$ in them will take on negative rather than positive values. We present numerical results of exact, self-consistent solution of the Friedmann equation and the phantom field equation of motion for several typical potentials in the Appendix (\S10.1).  These bear out that for typical values of $\delta w \lesssim 5\%$, the field will be in slow-roll and our formula $\delta w\propto H^{-2}$ will fit reasonably well.

\section{Techniques for observing DE}
\label{sec:observing}

Since many proposed experiments (e.g. WFIRST, Euclid, Planck) will use multiple methods to study DE, in this section we briefly review the physics and status of the main techniques, moving in the next to a discussion of specific experiments. For an up to date and extremely comprehensive review, see Weinberg et al. 2012; here our treatment will seek simply to provide basics sufficient to outline available tests for the theoretical results we have thus far developed.
\subsection{BAO}
Baryon Acoustic Oscillations (BAO) are one of the most promising ways to measure the expansion history of the universe (both the angular diameter distance $d_A$ and the Hubble constant $H$) and hence $w$ for DE.  Eisenstein et al. initially detected the BAO signature in 2005 with measurements of luminous red galaxies (LRGs) in Sloan Digital Sky Survey (SDSS-III) data, and since then the precision of this technique has significantly increased (Eisenstein et al. 2005, Seo \& Eisenstein 2007).  

The mechanism that produces the BAO is as follows.  In the hot, dense, early universe,  the photons couple to the baryons via Thomson scattering, so BAO, i.e.  sound waves in the primordial plasma, propagate and produce over and under-densities of baryons. Recombination at $z\sim 1100$ decouples the photons and baryons and ends the BAO. Thus the frozen-out wave is imprinted on the power spectrum we observe today with a characteristic comoving scale, providing a standard ruler to measure the subsequent expansion history (Eisenstein et al. 2005, Eisenstein 2011).  In a flat cosmological constant cosmology with parameters near the concordance model this co-moving scale is

\begin{equation}
r_s = 144.4\; {\rm Mpc} \left(\frac{\Omega_b h^2}{.24}\right)^{-.252}\left(\frac{\Omega_m h^2}{.14}\right)^{-.083}, 
\label{e:r_s}
\end{equation}
with $h \equiv H_0/(100\; {\rm km/s/Mpc})$, $\Omega_b$ the current density of baryons divided by the critical density, and $\Omega_m$ the current matter density divided by the critical density (cf. Zunckel et al. 2011). 

Planck will be able to deduce the value of $\Omega_mh^2$ to an accuracy of $1.25\%$ by studying the relative amplitudes of the baryon oscillation peaks in the CMB, and by their even and odd behavior in amplitude deduce the value of $\Omega_bh^2$ to an accuracy of $.64\%$ (see Zunckel et al. 2011 and Colombo et al. 2009).  Adding these errors in quadrature in eqn. (\ref{e:r_s}) yields a fractional error in $r_s$ of $.193\%$.  Planck will be able to measure the angular scale $\theta_s$ of the baryon oscillation peaks on the CMB sky to an accuracy of $.0519\%$.  Now
\begin{equation}
\theta_s = \frac{r_s}{d_{\rm{A}}(z=1089)}.
\label{e:baoscale}
\end{equation}

where $d_{\rm{A}}(z = 1089)$ is the angular diameter distance to the surface of last scattering at $z = 1089$.
Thus given the uncertainties in $r_s$ and $\theta_s$ , $d_{\rm{A}}(z = 1089)$ can be measured with an accuracy of $.200\%$.  This is an extraordinarily accurate measurement.  The BAO measurements using Eisenstein's technique and those using Park's topology technique can be normalized to this value (Park \& Kim 2009).   For instance, BOSS, currently underway, should measure the BAO scale to determine e.g. $d_{\rm{A}}(z =0.6)$  to an accuracy of $1.1\%$.

The major challenge for BAO measurements is non-linear structure formation, which smears out the peak in the galaxy-galaxy pair correlation function by jostling galaxies by $3-10\; \rm{Mpc}$.    Fortunately, since this effect is random and not systematic, it only leads to an order $.5\%$ systematic error which can be corrected for with the help of large N-body simulations.  Further, the effects of non-linear growth are not severe because the BAO scale has only gone mildly non-linear by now (Eisenstein et al. 2007, Eisenstein 2011).  For further discussion, see Padmanabhan \& White 2009, Noh et al. 2009, and Ivezic et al. 2011.

\subsection{Supernovae}

Type Ia supernovae (SNIa) are useful for measuring the luminosity distance $d_L$ because they are fairly standard candles: they all have roughly the same intrinsic luminosity. Hence measuring flux $F$ implies the luminosity distance $d_L$ via
\begin{equation}
F=\frac{L}{4\pi d_L^2}.
\end{equation}
It is not surprising that SNIa are roughly standard candles since they are all thought to be produced by the same physical process. When a white dwarf accreting mass from a binary companion reaches the Chandrasekhar mass limit, $M\sim 1.4 M_{\odot}$, the electron degeneracy pressure that supports it can no longer counterbalance gravity and the star explodes (Chandrasekhar 1931).

SNe measurements have several sources of error.  In order from the source to us, they are: evolution of supernovae with cosmological time, dust in supernovae's host galaxies, gravitational lensing, intersection with Earth's atmosphere (not a problem for space-based telescopes), the telescope itself, the filters, the detector response, and finally, the interpretation of the data (Perlmutter 2011).  For further discussion, see Ivezic et al. 2011; here, we touch briefly on dust and interpretation of data, as both are areas of recent progress.

Dust absorbs and scatters SNIa more in the blue than the red, and makes it hard to tell their intrinsic brightness. Recent work by Chotard et al. (2011) improves correction by using silicon (Si) and calcium (Ca) features to derive a dust-reddening law. They find a reddening law compatible with a Cardelli extinction law. This is a useful result because Cardelli extinction is well-understood, as it is what applies locally in the interstellar medium (ISM) (Cardelli et al. 1989).

Interpretation of data is complicated by the fact that SNe are not all exactly the same.  To make them better standard candles, ``stretch correction'' is applied. In 1992, Phillips observed that there is a correlation between the intrinsic brightness at maximum light and the duration of the light curve for SNIa (Li et al. 2011, Phillips 1993). This has been used to reduce the 1$\sigma$ spread in peak B-band luminosity from $0.3$ mag to $0.10 - 0.15$ mag (Kim 2004). Once this is done, further standardization is achieved by, at a given redshift, averaging over many supernovae and then comparing to an average over many supernovae at another, different redshift. More precise would be to compare supernovae that have exactly the same spectral features at different redshifts one-on-one. Efforts to do higher-precision distance measurements using these so-called ``supernovae twins'' are just beginning, but with promising results. So far, of 59 supernovae studied, 15 twins have been found. It is hoped that there are not many intrinsically different sub-types of SNIa, and hence that many twins will be found out of the supernovae for which spectra are already available (Perlmutter 2011).

We conclude by briefly noting what data sets are available for supernovae. The largest and latest (2010) data set, Union2, has 557 SNIa (Li et al. 2011, Perlmutter 2011). Other earlier samples include Union (307 SNIa, 2008) and Constitution (2009; added an additional 90 low-z SNe to Union).

\subsection{CMB, Weak Lensing, Clusters, \& Topology}
The temperature anisotropies in the cosmic microwave background (CMB) are affected by DE, and so measuring them constrains it.  The relative temperature anisotropies $\delta T/T$ are expanded in terms of the spherical harmonics $Y_{lm}$, and the power spectrum is therefore plotted as a function of the wavenumber $l$.  Hence the amplitude of the power spectrum at a given $l$ corresponds to the anisotropy on that angular scale.  DE alters the angular diameter distance $d_A$ and thereby changes the angular scale (and so the wavenumber) at which a given anisotropy occurs.  See Copeland et al. 2006 and Melchiorri et al. 2003 for further detail.  DE also alters the CMB through the Integrated Sachs-Wolfe (ISW) effect, a redshift that occurs when the gravitational potential $\Phi$ is time-dependent.  For $\Omega_m < 1$,  $\Phi$ varies with time, leading to an ISW effect that is particularly strong for large-scale power ($l \lesssim 20$) (Copeland et al. 2006).  

Finally, the low-$l$ modes of the CMB are most sensitive to DE, as they have most recently re-entered the causal horizon, and it is only at $z\lesssim 2$ that DE became dominant. Zinn (2012) shows that different values of the equation of state, specifically modeled as that of slow-roll DE, will shift the amplitudes of the modes with $l \lesssim 200$ (see Figure 1).  This is an additional avenue for CMB measurements to constrain the equation of state, especially expected to be fruitful because Planck should soon provide cosmic-variance limited measurements of these low-$l$ modes. Unfortunately, as Figure 2 illustrates, Zinn finds that  the low-$l$ modes of the CMB do not provide a very strong constraint at all on $\delta w_0$.  For further discussion of the method, see Zinn 2012.

\begin{figure}    
   
\includegraphics[scale=.25]{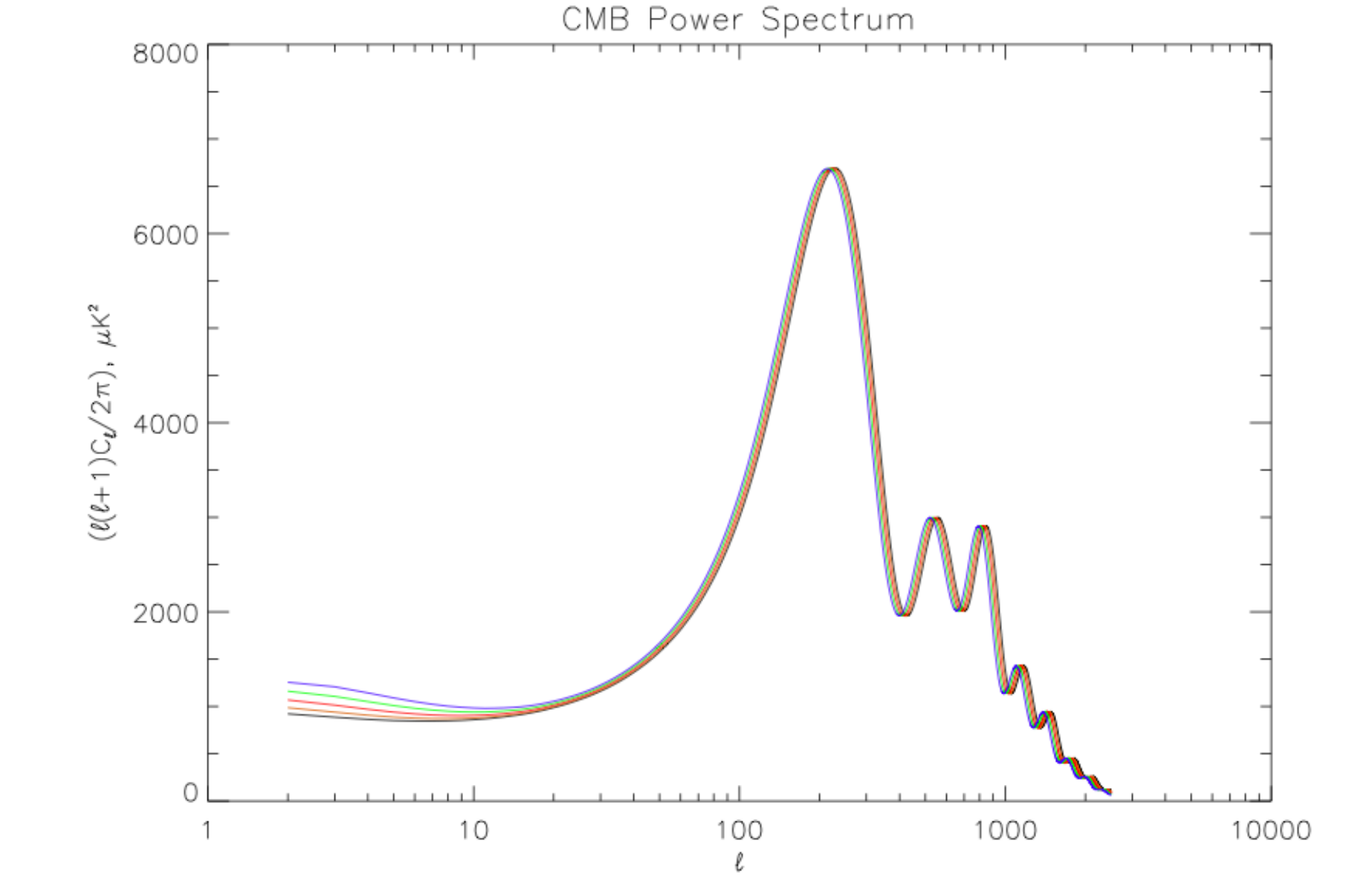} 

\centering{}\caption{Shift in amplitudes of low-$l$ modes for different values of $\delta w_0\equiv w_0 +1$.  Black, orange, and red correspond to phantom DE with values of $\delta w_0 = -.385,\;,-.105$ and $-.086$, respectively; green and blue correspond to scalar field DE with $\delta w_0 =. 173$ and $.453$. The high $l$ modes are shifted from right to left due simply to the effect of slow-roll DE on $d_A(z=1089)$.}

 \vspace{0.5in}
 
\label{f:ZinnCMB}

\end{figure}

\begin{figure}       

\includegraphics[scale=.25]{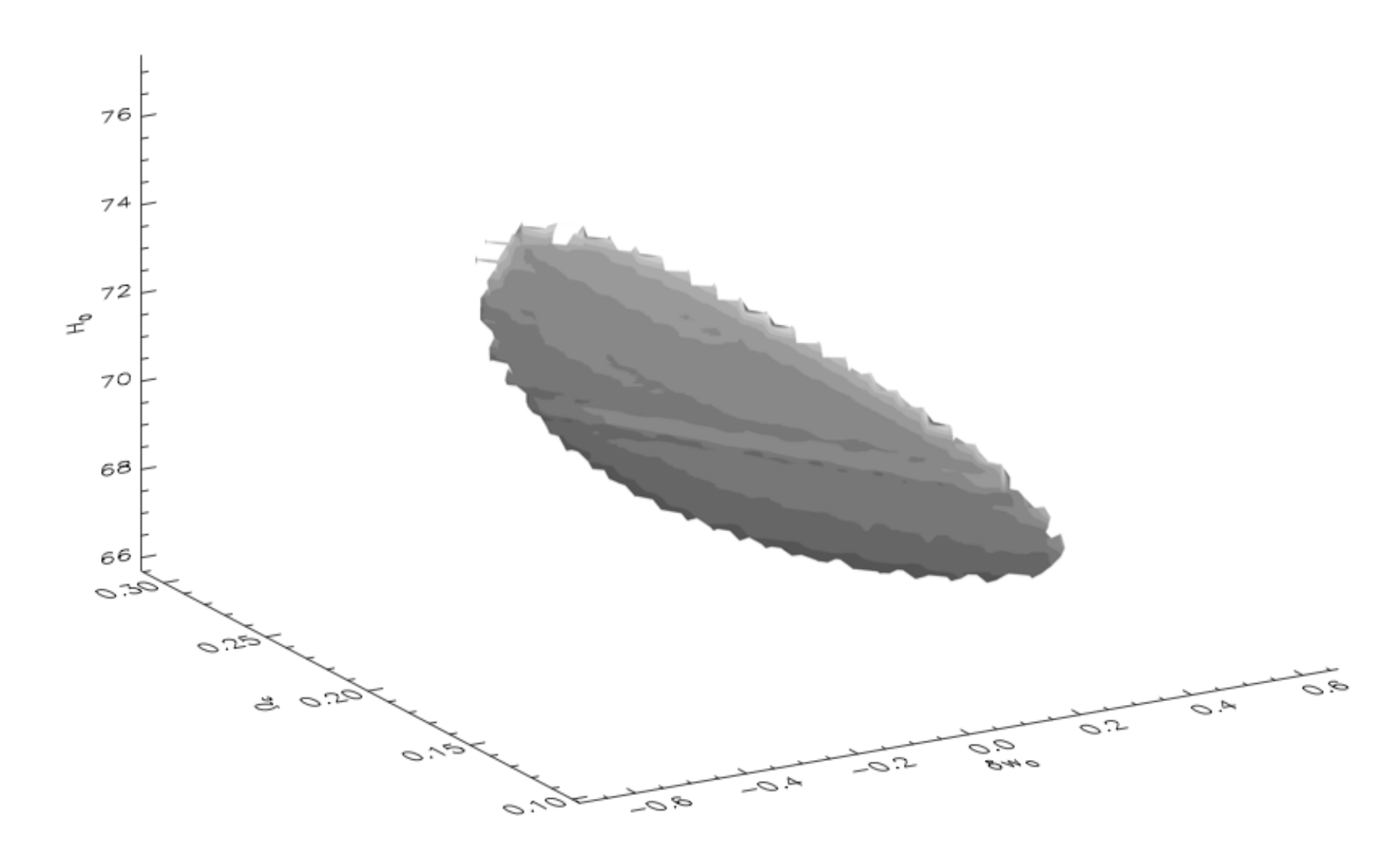} 

\centering{}\caption{A 95\% confidence ellipsoid showing the constraint low-$l$ CMB measurements offer between $H_0$ (in ${\rm km/s/Mpc}$) (vertical axis), $\Omega_m$ (left-hand axis), and $\delta w_0$ (rightmost axis).}

 \vspace{0.5in}
 
\label{f:Zinnellipsoid}
\end{figure}

Weak lensing (WL) is the distortion of the images of distant galaxies as the light we observe from them is bent by intervening matter.  First found in the '90's around individual halos, and detected in 2000 due to large-scale structure, weak lensing is a probe of the matter distribution and hence the role of DE in the growth of structure.  Lensing can create distortions in shape, size, and brightness; most easily measurable are distortions in shape, termed ``cosmic shear,'' which are $\simeq 1\%$.  Because typical intrinsic differences in galaxy shapes are $\simeq30-40\%$, a large sample of galaxies must be averaged over to detect weak lensing. For further discussion, see Li et al. (2011), Ivezic et al. (2011, and references therein), and Heavens (2009).

Lensing is also used to detect galaxy clusters (CL), which in turn can be counted as a function of mass and redshift and compared with simulation to shed light on DE's role in cosmological expansion.  Clusters can also be detected via optical and infrared imaging and spectroscopy, X-ray imaging and spectrocsopy, and the Sunyaev-Zel'dovich (SZ) effect.  For further discussion, see Li et al. 2011 and sources therein.

Additional constraints on DE can come from the topology of large-scale structure, for instance as measured using LRGs in SDSS-III (Gott et al. 2009).  It is hoped topology will provide independent measures of $w$ with about three-fifths the accuracy of the BAO method. Genus topology counts features in the cosmic web and thereby measures $r_s$ (Park \& Kim 2009).  The genus (number of donut holes minus number of isolated voids or clusters) per unit smoothing length cubed tells the physical scale because the genus depends only on the smoothed power spectrum at recombination.  This latter depends directly on $r_s$ and can be measured from the CMB.  Adopting smoothing lengths $l_s=(16 h^{-1}\;\rm{Mpc},\; 24 h^{-1}\;\rm{Mpc},\;34 h^{-1}\;\rm{Mpc})$, the genus as a function of volume fraction may be calculated as in Gott et al. 2009.  The genus per smoothing length cubed in each case provides the ratio $l_s/r_s$, with $r_s$ the BAO scale as in eqn. (\ref{e:baoscale}).  

The topology technique makes it possible to make an independent estimate of, for instance, $d_{\rm{A}}(z = 0.6)/ r_s$ (and therefore of $d_{\rm{A}}(z = 0.6)$) to an accuracy of $1.7\%$ using BOSS (Park et al. 2012, Speare 2012).  This estimate is independent of the estimate of $d_{\rm{A}}(z = 0.6)$ made by the BAO method because that method is using scales of $144\;\rm{Mpc}$ to fit the baryon oscillation features in the power spectrum while the topology method uses smoothing scales of $16 h^{-1}\;\rm{Mpc}$ to $34 h^{-1}\;\rm{Mpc}$ (which are also in the linear regime) to effectively fit the entire power spectrum.  Park and Kim 2009 have shown using N-body simulations that the genus measurement is particularly unaffected by non-linear and biasing effects.  The median density contour, which is what primarily determines the genus, does not change much as long as the smoothing length $l_s$ is large enough to put one into the linear regime.  This condition is met by the proposed smoothing lengths for LRGs.  If, for instance, the topology technique is used in addition to BAO and the errors are combined in quadrature, BOSS can measure e.g. $d_{\rm{A}}(z = 0.6)$ to an accuracy of $0.9\%$.

\section{Current and upcoming observations}
\label{subsec:currentandupcoming}
Numerous DE experiments are either in progress or upcoming in the next decade.  We treat the missions in two sections: those that have already at least begun physical construction, and those that have not.  We summarize all the missions we discuss (starting date and techniques) in Table \ref{t:surveys}.

\subsection{Current}

The Hubble Space Telescope (HST) made SNe measurements critical to confirming DE's existence, and it still is active in DE measurements, continuing its SNe work as well as being first to use the lensing around a galaxy cluster (CL) (Abell 1689) for DE (HST website).  
The Canada-France-Hawaii legacy survey (CFHTLS), which ran from 2003 to 2009 and has one data release still remaining, used SNe, via a deep survey to detect and monitor $\sim 500$ SNe Ia, and WL, via a wide survey over 170 square degrees.  It also made galaxy distribution measurements (using a wide survey) likely to be useful for topology (CFHTLS website).  In 2009, the Panoramic Survey Telescope and Rapid Response System (Pan-STARRS) began observing with its prototype single-mirror telescope; it will constrain DE via supernovae (SNe), weak lensing (WL), and clusters (CL) (Pan-STARRS website).  Also using clusters, detected via the Sunyaev-Zel'dovich (SZ) effect, the South Pole Telescope (SPT) will survey 4,000 square degrees and has already claimed the first use of clusters for cosmology (SPT website, Vanderlinde et al. 2010).

Planck, launched in 2009 and already reporting data, will use both CMB and CL to constrain DE. For further details, see Planck: the scientific programme, 2005,  Geisbuesch \& Hobson 2006, and de Putter et al. 2009.

The most significant near-term project using BAO is the Baryon Oscillation Spectroscopic Survey (BOSS).  Running from 2008 - 2014, it will measure $1.5$ million luminous red galaxies (LRGs) to $z=.8$.  This will constitute a seven-fold improvement  on the large-scale structure (LSS) data from SDSS-II: this comes from, first, a factor of two improvement in the instrument and, second, the fact that BOSS will focus on large-scale structure using more luminous galaxies that can be traced to larger distances (Eisenstein 2011, Eisenstein et al. 2011).   The next survey to use BAO will be the Large Sky Area Multi-Object Fiber Spectroscopic Telescope (LAMOST), a ground-based wide-field optical telescope on which construction was finished in 2008. Though limited by the small number of clear nights at its location in China (Hebei Province), it will produce the most accurate map to date of the baryons and dark matter in the Milky Way, constraining both DE and the growth of structure (LAMOST website, Day 2010).

Dark Energy Survey (DES), begun in late 2011, is one of the few surveys to use SNe, WL, CL, and BAO.  Since SNe and BAO measure the expansion rate, while weak lensing and clusters also measure the growth of structure, cross-comparison of these results can test not only DE but general relativity (DES website).  Hobby-Eberly Telescope Dark Energy Experiment (HETDEX), using the existing Hobby-Eberly Telescope at McDonald Observatory (Davis Mountains, Texas), began in January 2012 and is doing a three-year redshift survey of nearly one million galaxies to measure BAO (HETDEX website). 

\begin{table} 
 \vspace{0.2in}
\caption{Current and upcoming surveys: starting date and techniques used. Note that DES and LSST will yield $H(z)$ from the BAO, while BOSS, eBOSS, and BigBOSS will yield both $H(z)$ and $d_A$ (Blanton, personal comm.).}
 \vspace{0.04in}
\centering

\begin{tabular}{| l | c | c | c | c | c|}
    \hline
    Survey & Year  & SNe &  WL & CL & BAO \\ \hline
    HST & ongoing& X & X & X  & - \\ \hline
    CFHTLS & fin. 2009 & X & X & -  & - \\ \hline
    BOSS & begun & - & - & - & X \\ \hline
    SPT & first results & - & - & X & - \\ \hline
    Planck  & first results & - & - & X & - \\ \hline
    PanSTARRS & begun & - & X & X & - \\ \hline
    LAMOST & begun & ? & ? & ? & X  \\ \hline
    DES & 2012 & X & X & X & X \\ \hline
    HETDEX & 2012 & - & - & - & X \\ \hline
    BigBOSS & 2017 &  - & - & - & X \\ \hline
    eBOSS & 2014 & - & - & -  & X \\ \hline
    Euclid & 2017 &  - & X & - & X \\ \hline
    WFIRST & 2020 &   X  & X & - & X \\ \hline
    LSST & 2020 & X & X & - & X \\ \hline
    SKA & 2020 & - &  X & X & X \\ \hline
    ALPACA & ? & X & X & X & - \\ \hline
    
    \end{tabular}
     \vspace{0.2in}
     \label{t:surveys}
\end{table}

\subsection{Upcoming}

The Big Baryon Oscillation Spectroscopic Survey (BigBOSS), expected to begin surveying in 2017, will use luminous red galaxies (LRGs) to $z=1.0$ and bright OII emission line galaxies (ELGs) out to $z=1.7$ (20 million in total) to measure BAO to $.4\%$ for $.5<z<1.0$ and to $.6\%$ for $1<z<1.7$ (BigBOSS website).  Euclid (after the Greek geometer) is a satellite of the European Space Agency (ESA) with launch planned for 2017.  It will measure both WL and BAO, covering approximately half the sky in a 20,000 square degree survey; it also will have a 40 square degree deep survey mode (Euclid website; Refregier et al. 2010).  

Wide Field Infrared Survey Telescope (WFIRST) is a planned United States effort for 2020, though its funding prospects are uncertain.  It will use SNe, WL, and BAO, and should provide strong constraints on DE (Green et al. 2012, WFIRST website).  Ultimately the WFIRST mission goals may be accomplished using one of two Hubble Space Telescope-sized telescopes donated to NASA by the Department of Defense.  The Large Synoptic Survey Telescope (LSST), currently in the design and development and ``private construction'' phase, will use SNe, WL, and BAO, obtaining sub-percent precision in $H(z)$ and percent precision on the angular diameter distance at ten logarithmically-spaced redshifts (LSST website, Ivezic et al. 2011).  Square Kilometer Array (SKA) is an Australian effort to build a very-long baseline radio telescope array; in early 2012 it was decided that the telescope will have sites in South Africa, Australia, and New Zealand. It will survey a billion galaxies out to $z\simeq 1.5$, allowing use of the WL, CL, and BAO techniques (Blake et al. 2005).  Finally, Advanced Liquid-mirror Probe for Astrophysics, Cosmology, and Asteroids (ALPACA) will use SNe, WL, and CL, finding ~50,000 Type Ia SNe per year to $z\sim.8$ and $70,000$ galaxy clusters (ALPACA website).  Efforts  to gain funding for ALPACA are still active and a paper updating the science case is expected in 2013 (Arlin Crotts, personal communication).

\section{Method}
\label{sec:detection}

\subsection{Observational signature}
\label{subsec:signature}

We first compute the angular diameter distance $d_A$ and the luminosity distance $d_L$ for a cosmological constant cosmology by numerically evaluating eqns. (\ref{e:defdL}), (\ref{e:defdA}), and (\ref{e:comovdef}) (see \S10.3) using the Hubble constant as a function of redshift for a flat cosmology with negligible radiation and cosmological constant DE.  Note again this will only be valid for $z \ll 3196$, but since no observations take place at nearly such a high redshift, this is not a problematic restriction. We then compute $d_A$ and $d_L$ for a slow-roll DE cosmology by numerically evaluating the same equations ((\ref{e:defdL}), (\ref{e:defdA}), and (\ref{e:comovdef})) but now using the Hubble constant in a slow-roll cosmology, given by eqn. (14), rather than the Hubble constant in a cosmological constant cosmology.  We finally compute the fractional difference in luminosity and angular diameter distance between the slow-roll model and the cosmological constant model, $\Delta d_L/d_{L,-1}=\Delta d_A/d_{A,-1}=\Delta d_c/d_{c,-1}$ via eqn. (\ref{e:comoveq}), where the final equality involves the comoving distance (see Appendix \S10.3 for proof).  $\Delta d_L\equiv d_{L,SR}-d_{L,-1}$, with $d_{L,SR}$ the luminosity distance in a slow-roll cosmology and $d_{L,-1}$ that for a cosmological constant cosmology.  Analogous definitions are made for $\Delta d_{A},\;d_{A,SR}$, $d_{A,-1}$, and $\Delta d_c$ and $d_{c,-1}$.  

Our results are displayed in Figure 3---they are the observational signature of slow-roll DE.  This figure shows that the best $z$ to observe to see a difference between the two cosmologies in $d_A$ is $z\simeq 1$; notably, roughly where supernova searches are indeed able to measure $d_A$.   The best $z$ to observe to see how the cosmologies differ in $H$ is $z\simeq.5$---roughly where e.g. BOSS and BigBoss will look.  The blue curve (for $\Delta d_A/d_{A,-1}$) begins at zero because $\Delta d_{A,\;0}/d_{A,-1,\;0}=0=\Delta d_{L,\;0}/d_{L,-1,\;0}$ today---there is no difference between the slow-roll and cosmological constant cosmologies today because we normalize each to the same (current best) value of $H_0$ (thus far, we are neglecting errors in the cosmological parameters). As $z$ increases, DE influences $H(z)$, causing it to differ between slow-roll and cosmological constant cosmologies, and hence causing a difference in $d_c$ between the two.  This difference in $d_c$ is nearly monotonic in $z$ for small $z$ because $d_c$ stems from an integral over $z$; hence the slight difference at each redshift between the two cosmologies accumulates as one looks farther back into the past.  

However, at $z\simeq 1.5$, DE begins to be subdominant to matter, and since the two cosmologies being compared do not differ in the matter sector, the  fractional difference in comoving distance between the two, $\Delta d_c/d_{c,-1}$, begins to drop as the effects of DE become less and less important while the denominator $d_c$ continues to grow.  Note that the difference between the slow-roll DE cosmology's comoving distance and the cosmological constant cosmology's comoving distance is negative because the comoving distance is the physical distance with cosmological expansion ``taken out''. As eqn. (\ref{eq:smallzdH}) shows, a slow-roll DE cosmology will have expanded more in the past than a cosmological constant cosmology, so more expansion gets ``taken out'' to compute the comoving distance and hence the comoving distance in a slow-roll DE cosmology is lower than that in a cosmological constant cosmology.

The blue curve in Figure 3 for $\Delta d_c/d_{c,-1}$ fits with our small-$z$ approximate expression (see \S\ref{subsec:approxdadl} for derivation) which for low $z$ grows in magnitude with $z$ as 
 \begin{equation}
\big|\Delta d_c/d_{c,-1} \big| \approx \frac {3\Omega_{DE}\delta w_0}{4}z\;\;\;\;(small\;z).
 \label{eq:smallzdl}
 \end{equation}
 Computing the numerical value gives the slope of the blue curve in Figure 3 correctly for $z \ll  1$.
 
Now, we turn to the red curve, describing the fractional difference in the Hubble constant, $\Delta H/H_{-1} \equiv \left(H_{SR}-H_{-1}\right)/H_{-1}$, with all quantities defined analogously to those for $d_L$.  $\Delta H/H_{-1}$ also begins at zero, again because we have normalized both cosmologies to have the correct value of $H_0$ today (thus far, we are neglecting errors in the cosmological parameters).  It is monotonic in $z$ up to $z\simeq.5$; this is simply because the Hubble constant grows with increasing $z$, and for a slow-roll cosmology, the DE term will also have a $z$-dependence (see eqn. (\ref{e:slowrollfman})).  Hence the fractional difference between a slow-roll and a cosmological constant cosmology will grow with $z$ as long as DE is dominant.  

Just as with the curve for $\Delta d_c/d_{c,-1}$, the fractional difference begins to shrink when DE becomes subdominant to matter.  However, one sees this effect earlier in $\Delta H/H_{-1}$ than in $\Delta d_c/d_{c,-1}$.  This is because $\Delta H/H_{-1}$ depends more sensitively on the balance between matter and DE than does $\Delta d_c/d_{c,-1}$ because $\Delta H/H_{-1}$ is not an integrated difference whereas the latter is.  The integration from the present out to $z$ in $\Delta d_c/d_{c,-1}$ provides some ``cushion'' that masks the fact that DE is becoming sub-dominant to matter for a while, meaning $\Delta d_c/d_{c,-1}$ may continue to grow for a while even when $\Delta H/H_{-1}$ has already begun to turn around.

 The red curve in Figure 3 for $\Delta H/H_{-1}$ fits with our small-$z$ approximate expression as well (again, see \S\ref{subsec:approxdadl} for derivation) which for low $z$ grows with $z$ as 
 \begin{equation}
\Delta H/H_{-1} \approx\frac{3}{2}\Omega_{DE}\delta w_0 z\;\;\;\;(small\;z).
 \label{eq:smallzdH}
 \end{equation} 
 Computing the numerical value gives the slope of the red curve in Figure 3 correctly for $z \ll  1$.
 
Note that eqns. (\ref{eq:smallzdl}) and (\ref{eq:smallzdH}) of course break down when DE is no longer dominant.  At that point, as we have noted, the fractional differences $\Delta d_c/d_{c,-1}$ and $\Delta H/H_{-1}$ must turn around as the denominators (respectively, $d_{c,-1}$ and $H_{-1}$) continue to grow with $z$ while the numerators become ever-smaller as DE becomes subdominant.  The fact that $\delta w\propto \delta w_0 H^{-2}(z)$ increases this effect.  Thus, as the Hubble constant increases with $z$, since $\delta w \propto H^{-2}$, $\delta w\rightarrow 0$ as $z$ rises.  Therefore, even in the recent past, when DE is dominant, the slow-roll DE model tends towards the cosmological constant ($w\equiv -1$) as $z$ grows. Indeed, since $\delta w \propto H^{-2}(z)$ and the dynamical importance of DE falls off as $H^{-2}(z)$ as well (see eqn. (\ref{e:Happrx1})), the overall dynamical difference between the $\delta w_0 \neq 0$ (slow-roll) model and the $w\equiv -1$ model falls off like $H^{-4}(z)$.

\begin{figure}       

\includegraphics[scale=.8]{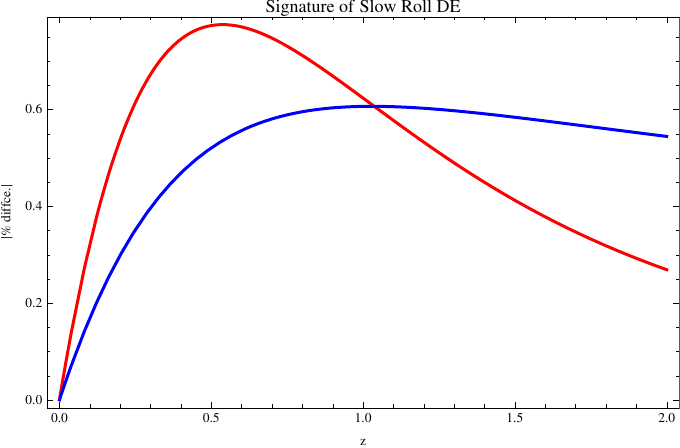} 

\centering{}\caption{The blue curve is the magnitude of the percent difference in comoving distance (which is the same as the percent difference for luminosity distance or angular diameter distance; see \S\ref{subsec:approxdadl}) between slow-roll and cosmological constant cosmologies as a function of redshift $z$.  We have plotted the magnitude because $\Delta d_c/d_{c,-1}$ is negative: the comoving distance in a slow-roll cosmology is lower than that for a cosmological constant cosmology. This is because a slow-roll cosmology will have expanded faster in the past than a cosmological constant cosmology, as eqn. (\ref{eq:smallzdH}) shows.  Of course the $\chi^2$ analysis we later carry out is insensitive to the sign of $\Delta d_c/d_{c,-1}$.  The red curve is the percent difference in the Hubble constant between the two cosmologies. Our calculations are for $\delta w_0=3.5\%$ as that is roughly consistent with the current best value of $w_0$ from WMAP+BAO+$H_0$.  The difference curves are zero today because we impose that the Hubble constant today has the same value in each model. The differences increase as we integrate back in time over $z$ and pick up the difference in the slow-roll versus cosmological constant equations of state.  The differences then turn around and move back towards zero because for $z\gtrsim 1$ DE is subdominant to matter and so the importance of a difference between cosmologies in the DE sector fades.}

 \vspace{0.5in}
 
\label{f:slowrollsig}
\end{figure}

\subsection{Ignoring cosmological parameter uncertainties}
\label{subsec:simplechi2}

Here, we use the observational signature discussed in \S5.1 to compute the possible confidence level of a detection of slow-roll DE as a function of $\delta w_0$.  For this first treatment, we ignore the uncertainties in the cosmological parameters and use the current value of $\Omega_{m}=.272$ from WMAP-7+BAO+$H_0$ (Komatsu et al. 2011).  

If the true cosmology is in fact a slow-roll DE one, at what confidence level might it be detectable? We compute the $\chi^2$  value and thence calculate a confidence level of the detection (see Fig. 4).  We use all the precisions available to us at the time of this calculation; these are as follows, and appear in tables in Appendix B (\S11).  We use WFIRST projected fractional precision on luminosity distance from SNe (optimistic) at 11 redshifts from $0.17$ to $1.15$; see Table 5.  We use WFIRST projected fractional precision on the Hubble constant at 13 redshifts from $0.8$ to $1.95$; see Table 6.  We use BigBOSS projected fractional precision on the Hubble constant at 16 redshifts from $0.15$ to $1.65$; see Table 7.  We use Euclid projected fractional precision on the Hubble constant at 12 redshifts from $0.7$ to $1.8$; see Table 8.  We use LSST projected fractional precision on the comoving distance at 9 redshifts from $0.5$ to $2.9$; see Table 9.  We use BOSS  projected fractional precision on the angular diameter distance at 3 redshifts from $0.35$ to $2.5$; see Table 10.  We use BOSS projected fractional precision on the Hubble constant at 3 redshifts from $0.35$ to $2.5$; see Table 11.   

In total, these precisions constitute $67$ degrees of freedom; an additional two are contributed by $H_0$ and $\Omega_m$ (see Table 12).  The $\chi^2$ is calculated by, at a given redshift, dividing the fractional difference between a slow-roll and cosmological constant cosmology by the fractional precision of any measurements at that redshift and adding the number of degrees of freedom; cf. eqn. (58), where measurements of angular diameter and luminosity distance will be incorporated in the same way as measurements of the comoving distance.  Assuming we have the correct values of $H_0$ and $\Omega_m$ means that the contribution from each of these to the total $\chi^2$ will be unity.

As a check on our numerical results, in Figure 5 we compare $\chi^2-DOF$, $DOF$ the number of degrees of freedom, with an analytical scaling (see \S10.4 for derivation):
\begin{equation}
\chi^2-DOF\propto \delta w_0^2
\label{eq:chi2wdw0}
\end{equation}
where the constant can be fixed by using $\chi^2$ for $\delta w_0=3.5\%$ as a fiducial value.

\begin{figure}       
\includegraphics[scale=.7]{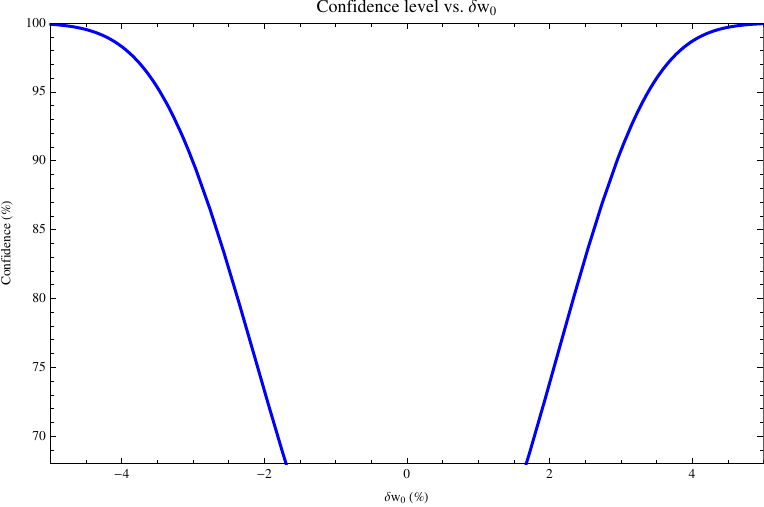} 
\centering{}\caption{The confidence level of a possible detection of slow-roll DE versus $\delta w_0\equiv w_0-1$, neglecting errors in the cosmological parameters. It is very encouraging to note that even if the deviation from a cosmological constant is only at the $2\%$ level, we might expect, in the limit of highly accurate cosmological parameters (specifically, as \S5 will discuss, the matter density and Hubble constant), a detection at approximately $75\%$ confidence.  If $\delta w_0 = 3.5\%$, the confidence of a detection would rise to $96\%$.}
 \vspace{0.5in}
\label{f:confidencenoerrs}
\end{figure}

\subsection{Incorporating cosmological parameter uncertainties}
\label{subsubsec:numresults}

There are a number of sources of error in any cosmology.  There will be error due to uncertainty in the matter density, the radiation density, the curvature, the DE density, and the value of the Hubble constant. We have already discussed the extent to which one might discriminate between slow-roll DE and a cosmological constant if future methods allowed us to
determine $\Omega_m$ and $H_0$ to high accuracy (\S5.2 and Figure 4).  We now consider two further cases:  i) if the main significant error in
cosmological parameters is a 1.25\% uncertainty in $\Omega_m$ (\S5.3.1 and Figs. 6 and 8), and ii) if we
also allow for uncertainty in $H_0$ (\S5.3.2 and \S6, Figs. 7, 9-11).  

In case ii), we simply add a 1\% uncertainty in $H_0$ from all non-DE experiments to our work. We expect this to be an overestimate. This is the simplest approach. Considering possible correlations between the errors in $H_0$ and $\Omega_m$, which for instance would occur if the measurement of $\Omega_m$ came from $\Omega_m h^2$ {\it and} the uncertainty in $h$ were not negligible compared to that in $\Omega_m$, is beyond the scope of this work, as the precisions that future experiments will be able to achieve on these parameters may only be estimated at present in any case. We do display the confidence level of a detection for different values of $\delta w_0$ as a function of the precision on $\Omega_m$, with 1\% precision on $H_0$ from all non-DE experiments, in the Appendix (\S10.3).  This is to illustrate that, for reasonable expected values of the precision on $\Omega_m$, even if they differ from 1.25\%, one can still expect a detection with good levels of confidence. 

We now turn to discussion of case i) in the following section and case ii)  in \S5.3.2.

\subsubsection{In current densities}

Our model is motivated by inflation, which implies a flat Universe. If inflation is persuasive, then the reader will believe that the Universe really is flat, meaning spatial curvature is actually zero to high accuracy.  This also means we are at the critical density today, so the densities of matter, radiation, and DE must sum to unity. The radiation density is $10^{-5}$ today, so we neglect it; this approximation will never be a significant source of error at the redshifts we consider, as we have noted earlier several times.  We therefore have the constraint that $\Omega_{DE}+\Omega_m=1$, which means that an overestimate of DE will correspond to an underestimate of matter, and vice versa.  Because this is so, {\it we need not directly consider errors in the DE density.} Should one desire the effects of an overestimate of $x\%$ in the DE density, one just needs to consider a corresponding underestimate of the matter density, given by $-2.7 x\%$, because $\Omega_{DE}\rightarrow(1+x)\Omega_{DE}$, so $\Omega_m\rightarrow(1-x\Omega_{DE}/\Omega_m)\Omega_m=(1-y)\Omega_m$, with $y=x(\Omega_{DE}/\Omega_m)\approx 2.7 x$.

We use a projected precision on $\Omega_m$ today of 1.25\% from Planck; see \S3.1.  This is the
uncertainty in $\Omega_m h^2$ which Planck will obtain.  If an
accurate value of $h\equiv H_0/(100 {\rm  km/s/Mpc})$ is obtained by other experiments, then we should know
$\Omega_m$ at least this accurately. 

For instance, we can determine $h$ to an accuracy of 1\%
from the Planck CMB data alone, with the flatness assumption. BOSS and BigBOSS alone, together with model
fitting of the dark matter model, should lead to a value of $h$ with a statistical
uncertainty of $0.3\%$. Given the $0.193\%$ uncertainty in $r_s$, this would give an uncertainty in $h$ of $0.48\%$ from BOSS and BigBOSS alone. Therefore, combining measurements of $h$ from the Planck CMB data and from BOSS and BigBOSS should yield an estimate of $h$ with
an accuracy of $0.43\%$.

Other improvements
in $h$ could come from supernova measurements, gravitational lensing
time delays, and gravity wave detections from LISA (if its original funding were restored) (Phinney 2002). Hence, we will assume improvements in $h$
measurement over the next seven years, using these methods and others, to be
sufficient that uncertainty in $h$ is not the limiting factor in determining $\Omega_m$.

Further, we note that $\Omega_m$ may be measured independently using
masses of clusters of galaxies calibrated by large N-body simulations.
$\Omega_m$ can also be measured using the amplitude of large-scale velocity
perturbations. These points bolster our claim that it is conservative to estimate that $\Omega_m$ may be measured to $1.25\%$. For additional discussion of the effects of errors in $\Omega_m$, see Alam et al. 2007 and Sahni et al. 2008.


\subsubsection{In the Hubble constant}

We now turn to the Hubble constant.  All BAO and topology measures of $d_{\rm{A}}(z)/ d_{\rm{A}}(z = 1089)$ are to be compared with models.  This is because $d_A(z=1089)$ can be determined to high accuracy ($0.2\%$) from the CMB. Likewise SNe measurements of $d_{\rm{A}}$ (from $d_{\rm{L}}$) can be compared with each other.  Distant supernovae can be compared with nearby ones to determine relative values of $d_{\rm{A}}(z)$ normalized by nearby supernovae. Thus, in both cases, uncertainty over the value of $H_0 = 73.8\pm 2.4\;\rm{km}\rm{s}^{-1}\;\rm{Mpc}^{-1}$ (Riess et al. 2011), which is currently at the $3.3\%$ level, can be eliminated.   

However, direct measurements of the Hubble constant, such as those to be done by WFIRST, Euclid, BOSS, and BigBOSS, are sensitive to uncertainty in the value of $H_0$.  These data points constitute $65\%$ by number of the points we use in our analysis, so we must incorporate the possible effects of an error in $H_0$. We take it that in the next five years, $H_0$ will be constrained to on order $1\%$ precision (Lyman Page, personal communication), and use this value to compute the $\chi^2$ penalty a value of $H_0$ different from the current best value will incur in our simulations.

\subsubsection{Method}
 
 Our concern is to understand the effects of an error in $H_0$ or in $\Omega_m$ on the observations.  If the true cosmology were slow-roll DE, but an error in $H_0$ or in $\Omega_m$ (or both) were made and a cosmological constant cosmology assumed, might the error mimic the effect of slow-roll DE and thus allow the (false) cosmological constant cosmology to fit the observations well?
 
We calculate the difference between observables, such as angular diameter distance and Hubble constant, for a $w\equiv -1$ cosmology with incorrect values of $H_0$ and $\Omega_m$ and a slow-roll DE cosmology with the correct $H_0$ and $\Omega_m$, which we take to be $\Omega_m=.272$ from the WMAP-7+BAO+$H_0$ value and $H_0 = 73.8\pm 2.4\;\rm{km}\rm{s}^{-1}\;\rm{Mpc}^{-1}$ from Riess et al. 2011. The precisions and experiments used are described in detail in \S\ref{subsec:simplechi2}, and summarized in tables in Appendix B (\S11).

For a given $\delta w_0$ and $H_0$, we sample a number of ``wrong" values of $\Omega_m$.  The results for several $\delta w_0$'s are displayed as solid curves in Figure 6.  The minima of the parabolas in that figure are the $\Omega_m$'s in a cosmological constant cosmology that will best ``mimic" slow-roll DE cosmologies with values of $\delta w_0$ as given on the horizontal axis and a correct value of $H_0$.  They will hence be the values that, given the correct $H_0$, make it most difficult to distinguish the two different cosmologies.  

We now momentarily restrict ourselves to errors in $\Omega_m$ only, and assume we have the correct value for $H_0$.  We discuss several numerical checks done on these results below, considering as a typical example the slow-roll case with $\delta w_0=3.5\%$.  Following this discussion, in \S5.3.7 we return to the main thread of \S5.3 and detail how uncertainty in $H_0$ is incorporated.

For $\delta w_0=3.5\%$, the ``mimic'' matter density, denoted with an additional subscript ``m'', is $\Omega_{mm}=.278392$, $2.55\%$ larger than the true value of $\Omega_m$.  If the true cosmology were $w\equiv-1$ but with this value of $\Omega_m$, there is a $27\%$ chance that we would mistakenly see observational results that mirrored a slow-roll DE cosmology with $\Omega_m=.272$ and $\delta w_0 = 3.5\%$.

Why is the mimic value of $\Omega_m$ larger than $.272$?  Slow-roll DE has a higher average value of $w$ than does cosmological constant DE. Having $w\equiv -1$ but with an over-estimate of matter relative to the true value has the effect of raising $w_{\rm{tot}}$, the average total ratio of pressure to energy density, because matter has $w=0$.  Slow-roll DE has $w>-1$, so it also has this effect.  Thus it is no surprise that extra matter can mimic slow-roll DE.  Too much extra matter, though, leads to a $\chi^2$ penalty into the past, as the SR DE cosmology diverges from the $w\equiv-1$ (with wrong $\Omega_m$) cosmology in the matter-dominated epoch when that wrong amount of matter becomes more dominant.

We now discuss three methods we use to check the plausibility of our numerical results. 

\subsubsection{Small-$z$ expansion}

 First, one may expand eqn. (\ref{e:slowrollfman}) for $z \ll 1$ with the binomial expansion; it is then linear in $z$. It can then be set equal to $H^2(z)$ for $w\equiv-1$.  One can then solve for the ``mimic" $\Omega_m$, denoted $\Omega_{mm}$.
\begin{eqnarray}
H^2_{\rm{SR}} & \approx &H_0^2 \left(1+3\left(\Omega_m+\Omega_{\rm{DE}}\delta w_0\right)z\right)\nonumber\\
&=& H^2_{-1}\approx H_0^2\left(1+3\Omega_{mm}z\right).
\end{eqnarray}
This yields 
\begin{equation}
\Omega_{mm}=\Omega_m+\Omega_{\rm{DE}}\delta w_0.
\label{e:mimicsmallz}
\end{equation}

\subsubsection{Average total equation of state today}
Second, one may define $\bar{w}_{\rm{tot}}=\sum\Omega_i w_i/\sum\Omega_i$ and ask for what $\Omega_{mm}$ is so that $\bar{w}_{\rm{tot},\;\rm{SR}}=\bar{w}_{\rm{tot},\;w\equiv-1}$, where a subscript ``tot'' denotes the total ratio of pressure to energy density.  In other words, at present, what $\Omega_{mm}$ is required in the $w\equiv-1$ cosmology so that the overall ratio of pressure to energy density is the same as that in the slow-roll DE cosmology? This leads to the same result as given in eqn. (\ref{e:mimicsmallz}). This is unsurprising as eqn. (\ref{e:mimicsmallz}) is from a small-$z$ expansion about $z=0$, the present.

Evaluating eqn. (\ref{e:mimicsmallz}) yields $\Omega_{mm}=.29748$.  From earlier in this section, the numerical $\Omega_{mm}=.278392$.  Recall that the latter value is obtained by finding the $\chi^2$ for all the proposed future observations as a function of $\Omega_m$ and seeing which value of $\Omega_m$ in a cosmological constant cosmology best mimics a slow-roll model with $\delta w_{0}=3.5\%$.  So we would expect these results to be very close to each other.  However, the analytical result from eqn. (\ref{e:mimicsmallz}) is farther above the true value of $\Omega_m=.272$ because the analytical analysis includes only small redshifts $z \ll 1$, whereas the numerical results include the past out to $z \simeq 2$.  Being wrong on $\Omega_m$ increases the $\chi^2$ more and more as one goes further back into the past, so we would expect that the numerical $\Omega_{mm}$ is closer to the true value of $\Omega_m$ than the analytical $\Omega_{mm}$ from eqn. (\ref{e:mimicsmallz}) is. To check that intuition, we compute a numerical $\Omega_{mm}$ by the methods described above but using only observations $z\leq1$.  We would expect this result to be higher than $\Omega_{mm}$ as computed using the full set of observations (i.e. including higher redshift observations) because it will not be paying the $\chi^2$ penalty of being more and more wrong farther back in the past, since points with $z>1$ are not included. We also expect it to be closer to $\Omega_{mm}$ as from eqn. (\ref{e:mimicsmallz}) because that equation is valid for small-$z$ and $\Omega_{mm}$ will now be computed using only small-$z$ ($z<1$). The value obtained is $\Omega_{mm}=.278853$, which fulfills these expectations.

Finally, we can even fit a line to the two points from the numerical analysis (all $z$, and $z<1$ only) as a function of $\bar{z}$, the average redshift of the observations used.  In other words, what is the line that fits both $(\bar{z}_{full},\Omega_{mm})$ and $(\bar{z}_{small},\Omega_{mm})$, where the former $\Omega_{mm}$ is from the full numerical $\chi^2$ with all observations as described earlier in this section and the latter is from the small-$z$ only subset as described in the previous paragraph. The line fitting these two points can then be used to extrapolate and find a prediction for the value of $\Omega_{mm}$ if only today ($z=0$) were taken into account.  This result is $\Omega_{mm,0}=.279397$, larger than both $\Omega_{mm}$ for small-$z$ and the numerical result for $\Omega_{mm}$, as expected, and closest of this set to the value from eqn. (\ref{e:mimicsmallz}).

\subsubsection{Comparison with analytical $\chi^2$ scaling}
Third, we can compare the numerical results giving $\chi^2$ for different values of $\Omega_m$ with an analytical scaling (see Appendix \S10.4 for derivation):

\begin{equation}
\chi_H^2-DOF \propto \left[1+\alpha \left(\Delta \Omega_m \right) \right]^2,
\label{e:chisqscalingwithmatter}
\end{equation}

where $\Delta \Omega_m$ is the difference between the value of the matter density used for the cosmological constant cosmology and the true value of $\Omega_m=.272$.  $DOF$ is simply the number of degrees of freedom. The subscript ``H'' indicates that this scaling is only strictly valid for the contribution to the $\chi^2$ from observations of $H$; it is very complex to derive an analytical scaling for the contributions that depend on the luminosity or angular diameter distances. See Figure 6 for a comparison of this scaling with the numerical results.

We have defined $\alpha=1/\left(A(2)\delta w_0\right)$, with 
\begin{equation}
A(u)\equiv \frac{1}{u^3-1}\ln \left[ \frac{u^3}{\Omega_m u^3+\Omega_{DE}} \right]
\end{equation}
and $u\equiv1+z$.
We evaluate $A(u)$ at $<u>=1+<z>$ with $<z>=1$, since that is roughly the average redshift of an observation used in computing the $\chi^2$ (see \S10.4).

\begin{figure}       
\includegraphics[scale=.7]{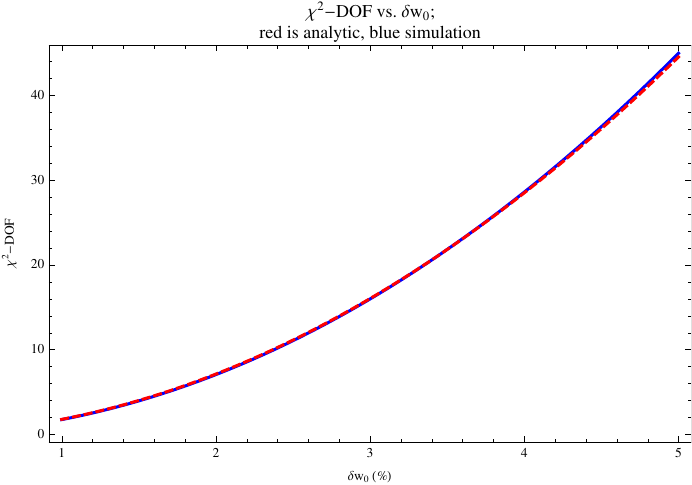} 
\centering{}\caption{The red dashed curve is the analytical formula eqn. (\ref{eq:chi2wdw0}), the blue solid curve the numerical results. The vertical axis is the $\chi^2$ minus the number of degrees of freedom (DOF).  This shows that our analytical scaling for $\chi^2$ is an extremely accurate predictor of the true numerical values. This plot assumes no error in the cosmological parameters.}
 \vspace{0.5in}
\label{f:chi2fnofdw0}
\end{figure}

\begin{figure}       
\includegraphics[scale=.7]{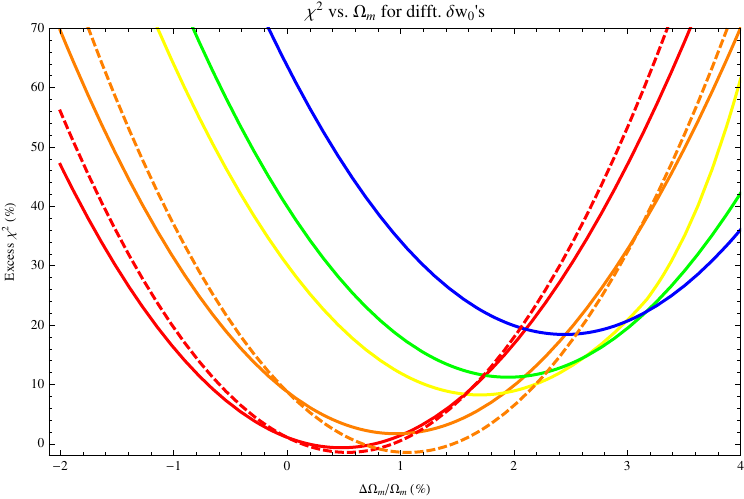} 
\centering{}\caption{Red is $\delta w_0=1\%$, orange $2\%$, yellow $3.5\%$, green $4\%$, and blue $5\%$.  Solid curves are numerical results, dashed curves are using analytical formula eqn. (\ref{e:chisqscalingwithmatter}). We have only plotted the analytical scaling for two curves so that the plot remains legible.}
 \vspace{0.5in}
\label{f:chi2vsdeltomega}
\end{figure}

Now, observations of $H$ represent $65\%$ of the observations we use, so we expect the scaling to be good but not perfect. It is both this point and the fact that we evaluate $A(u)$ at an average value of $u=1+z=2$ that cause the analytical formulae to be shifted slightly to the right of the numerical results in Figure 6. Indeed, one can calculate $\chi^2$ using only Hubble constant measurements to eliminate the former point; even then the analytical results do not perfectly mirror the numerical ones, showing that evaluating $A(u)$ at an average redshift does introduce some error.  However, the analytical results agree well enough with the numerical results to persuade that the latter are accurate.

\subsubsection{Adding uncertainty in $H_0$}

We iterate the process described in \S5.3.3 over different values of the Hubble constant $H_0$.  Ultimately, we thereby obtain a hyper-surface giving $\chi^2$ as a function of: $\delta w_0$, the error in the matter density $\Delta \Omega_m$, and the error in the Hubble constant $\Delta H_0$.  We show several representative  slices through this hyper-surface in \S6, Figures 8-10.  

To obtain the confidence with which slow-roll DE might be detected for a given $\delta w_0$, we calculate the value of $\Delta H_0$ and $\Delta \Omega_m$ for which $\chi^2$ is minimized.  In other words, we ask, for each value of $H_0$ and $\delta w_0$, what $\Omega_m$ minimizes the confidence of a detection. Then, we allow the value of $H_0$ to vary to minimize over this set.  This gives the minimum confidence of a detection for a particular $\delta w_0$.  Essentially, at a given $\delta w_0$, we look at many slices of constant $H_0$, for instance as given in Figures 8-10, and find the $\Omega_m$ lying on the lowest-confidence ellipse in each slice.  We then seek the slice of constant $H_0$ for which the confidence associated with this point will be least. This procedure yields the least favorable combination of $\Omega_m$ and $H_0$ for distinguishing slow-roll DE with a given $\delta w_0$ from a cosmological constant DE cosmology with different cosmological parameters.  The resulting confidences are plotted in Figure 11.

The checks described in \S5.3.4 and \S5.3.5 do not apply to uncertainty in $H_0$, since term-by-term equality in powers of $z$ in eqn. (28) is no longer possible if $H_0$ differs between the slow-roll and cosmological constant cosmologies.  However, the same type of check as that described in \S5.3.6 does apply; we derive a scaling formula (eqn. (\ref{e:chisqvsh})) in \S10.4, and compare it with the numerical results in Figure 7. The agreement should persuade that the numerical results are accurate.

\begin{figure}       
\includegraphics[scale=.7]{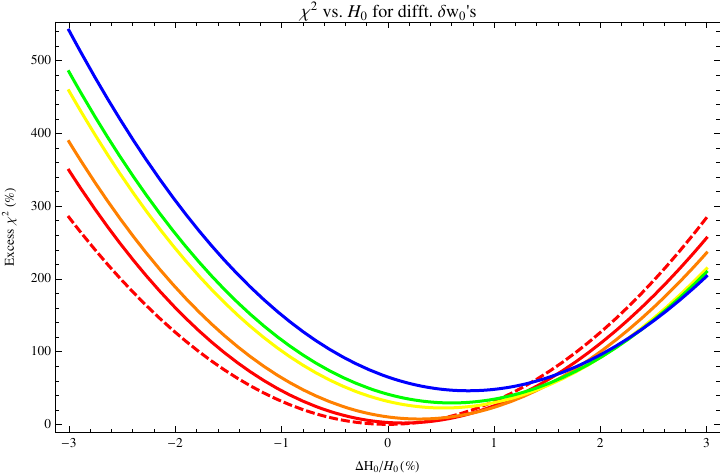} 
\centering{}\caption{Red is $\delta w_0=1\%$, orange $2\%$, yellow $3.5\%$, green $4\%$, and blue $5\%$.  Solid curves are numerical results, dashed curve is using analytical formula eqn. (\ref{e:chisqvsh}). We have only plotted the analytical scaling for one curve so that the plot remains legible.}
 \vspace{0.5in}
\label{f:chi2vsdeltah}
\end{figure}


\section{Results}
Using the method outlined in the previous section, we calculate the least favorable combination of errors $\Delta \Omega_m$ and $\Delta H_0$ for distinguishing a slow-roll DE cosmology from a cosmological constant cosmology for each $\delta w_0$ we study.  These are the errors $\Delta \Omega_m$ and $\Delta H_0$ in a cosmological constant cosmology that would best mimic a slow-roll DE cosmology with $\Omega_m=.272$ and with a certain value of $\delta w_0$. Hence this is the cosmological constant cosmology that would be hardest to distinguish from a slow-roll DE cosmology with $\Omega_m=.272$.  Thus it will produce the lowest $\chi^2$ and the lowest confidence level of detection.  

We hope the confidence levels our analysis yields will show the value of funding as many experiments as possible.  Given that this may not occur, however, we also provide several breakdowns of the $\chi^2$ contributions, which may be helpful in assessing which experiments will yield the greatest evidence for or against slow-roll DE.  We provide the individual $\chi^2$ for each experiment (Table \ref{t:individchisq}), bin them into space- and ground-based (Table \ref{t:spacevsground}), and bin them into experiments beginning in the next five years and experiments beginning six years from now or later (Table \ref{t:nextfivevssixplus}). In the first of these tables, we show the values obtained neglecting errors in the cosmological parameters as well as those obtained accounting for them.

Figures 8-10 show the confidence level resulting from the $\chi^2$ analysis versus $\delta w_0$. Figure 11 condenses these results into a single curve giving the confidence values as a function of $\delta w_0$ for the least favorable values of $\Delta \Omega_m$ and $\Delta H_0$.  What is encouraging is that, even assuming the worst-case scenario that our values of the matter density and $H_0$ are incorrect in precisely the way least favorable to distinguishing slow-roll from cosmological constant DE, for the value $\delta w_0=3.5\%$, we may still expect a detection at $70\%$ confidence using experiments taking place in the next seven or so years. Furthermore, should $\delta w_0$ be larger (e.g $5\%$), a possibility still very much observationally allowed, the confidence of a detection might rise to $\simeq 85\%$.

\begin{table}
 \vspace{0.2in}
\caption{$\chi^2$ for individual experiments if $\delta w_0=3.5\%$.  The confidence level is one minus the probability of getting $w\equiv -1$-like observations by chance if slow-roll DE is the true cosmology.  The third column assumes we know the cosmological parameters $\Omega_m$ and $H_0$ to arbitrary accuracy.  The final column, ``$\chi^2$ w/errs.'', is the $\chi^2$ value of a detection (and the bottom row the confidence of a detection) when the possibility of errors in the matter density $\Omega_m$ and the Hubble constant $H_0$ has been accounted for as described in \S5.  The value in the ``$\chi^2$ w/errs.'' column for ``Wrong $H_0$'' is actually $1.0034$, so we have accounted for the possibility of a wrong $H_0$ (and wrong $\Omega_m$); in contrast, in the ``$\chi^2$'' column, these parameters are both constrained to have the correct values.}
 \vspace{0.04in}
\centering
    \begin{tabular}{| l | c | c | c| }
 
    \hline
     Experiment & DOF & $\chi^2$ & $\chi^2$\; \rm{w/errs.}\\ \hline
     BOSS $d_A$ & 3 & 3.56 & 3.20  \\ \hline
     BOSS H & 3& 3.38 & 3.29  \\ \hline
     LSST BAO/WL & 9 & 19.88 & 10.67  \\ \hline
     WFIRST SNe & 11 &14.27 & 12.23  \\ \hline
     WFIRST H &  13 & 14.83 & 13.50 \\ \hline
      Euclid H &  12 & 13.26 & 12.17 \\ \hline
      BigBOSS H & 16 & 19.63 & 16.60\\ \hline
      Wrong matter & 1 & 1.00 & 2.99  \\ \hline
       Wrong $H_0$ & 1 & 1.00 & 1.00  \\ \hline
      Total & 69 & 90.81 & 75.65   \\ \hline
      Confidence &--& 95.96\%  & 72.75\% \\ \hline
      
      \end{tabular}
     \vspace{0.2in}
     \label{t:individchisq}
\end{table}

\begin{table}
 \vspace{0.2in}
\caption{Contribution to $\chi^2$ of ground-based or space-based observations each; with $\delta w_0 = 3.5\%$. We have split the matter and $H_0$ contributions equally between ground and space-based experiments, as each parameter will be constrained by both and a more sophisticated split would be unnecessarily complex.}
 \vspace{0.04in}
\centering
    \begin{tabular}{| l | c | }
 
    \hline
     Expmt. & \% of $\chi^2 \;\rm{w/errs.}$ \\ \hline
     Ground & 47.53\% \\ \hline
     Space & 52.74\% \\ \hline
     \end{tabular}
     \vspace{0.2in}
     \label{t:spacevsground}
\end{table}

\begin{table}
 \vspace{0.2in}
\caption{Contribution to $\chi^2$ of experiments in the next five years only or in six years plus; with $\delta w_0=3.5\%$. We have counted the matter and $H_0$ contributions as coming solely from observations in the next five years, as the precisions we have used for these are expected to be achieved in that timeframe.}
 \vspace{0.04in}
\centering
    \begin{tabular}{| l | c | }
 
    \hline
     Expmts. in  &  \% of $\chi^2 \;\rm{w/errs.}$\\ \hline
     Next five only & 51.88\% \\ \hline
     Six plus only & 48.12\%\\ \hline
     \end{tabular}
     \vspace{0.2in}
     \label{t:nextfivevssixplus}
\end{table}

\begin{figure}       
\includegraphics[scale=.6]{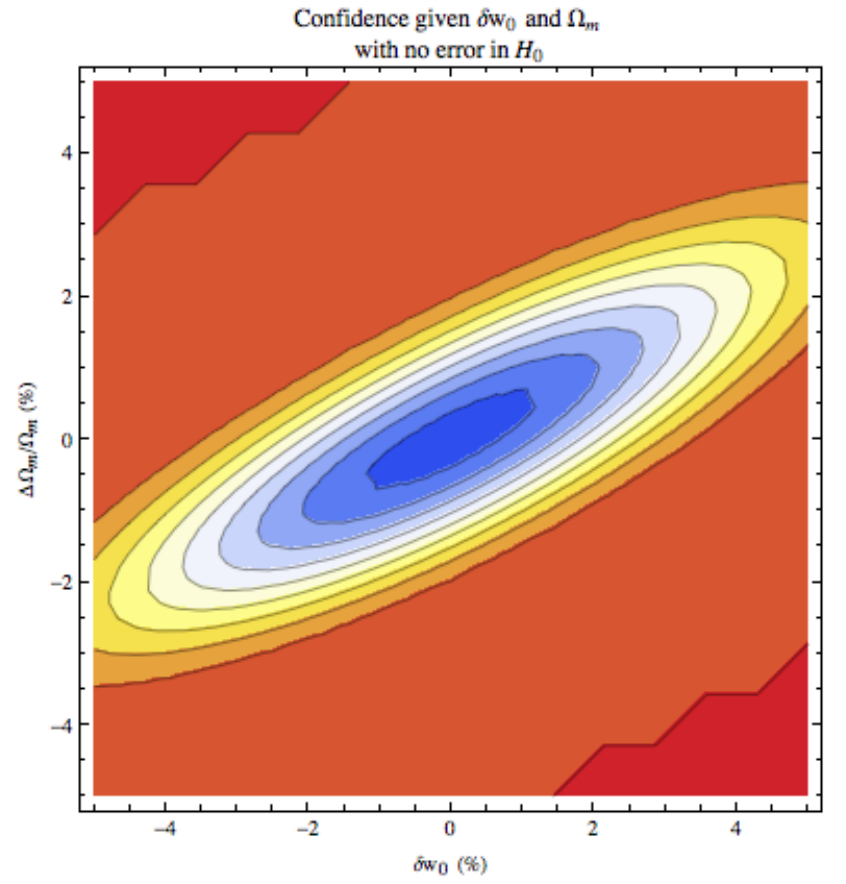} 
\centering{}\caption{The confidence of a detection as a function of $\delta w_0$ and $\Omega_m$ assuming no error in $H_0$. Warmer colors denote a higher-confidence detection of slow-roll DE; the scale should be read just like a thermometer. The orange ellipse intersecting $(-5,-3.5)$ is $95\%$ confidence; the white ellipse $75\%$, and the center blue $55\%$.}
 \vspace{0.5in}
\label{f:res1}
\end{figure}

\begin{figure}       
\includegraphics[scale=.6]{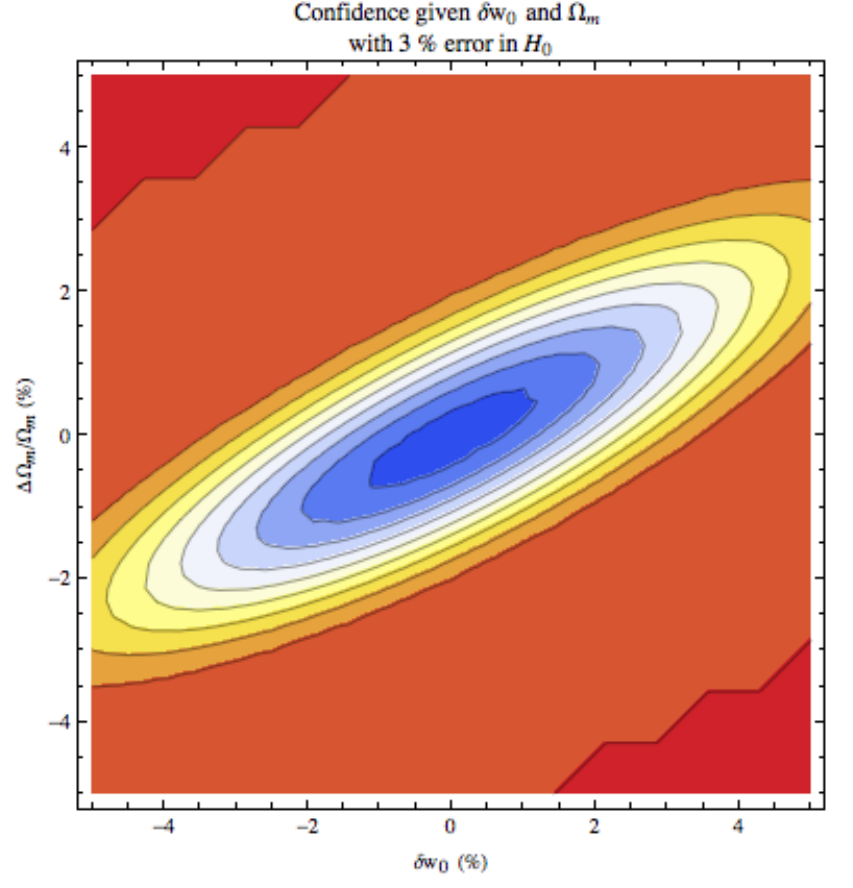} 
\centering{}\caption{The confidence of a detection as a function of $\delta w_0$ and $\Omega_m$ assuming a $3\%$ error in $H_0$. Warmer colors denote a higher-confidence detection of slow-roll DE; the scale should be read just like a thermometer. The orange ellipse intersecting $(-5,-3.5)$ is $95\%$ confidence; the white ellipse $75\%$, and the center blue $55\%$.}
 \vspace{0.5in}
\label{f:res2}
\end{figure}

\begin{figure}       
\includegraphics[scale=.6]{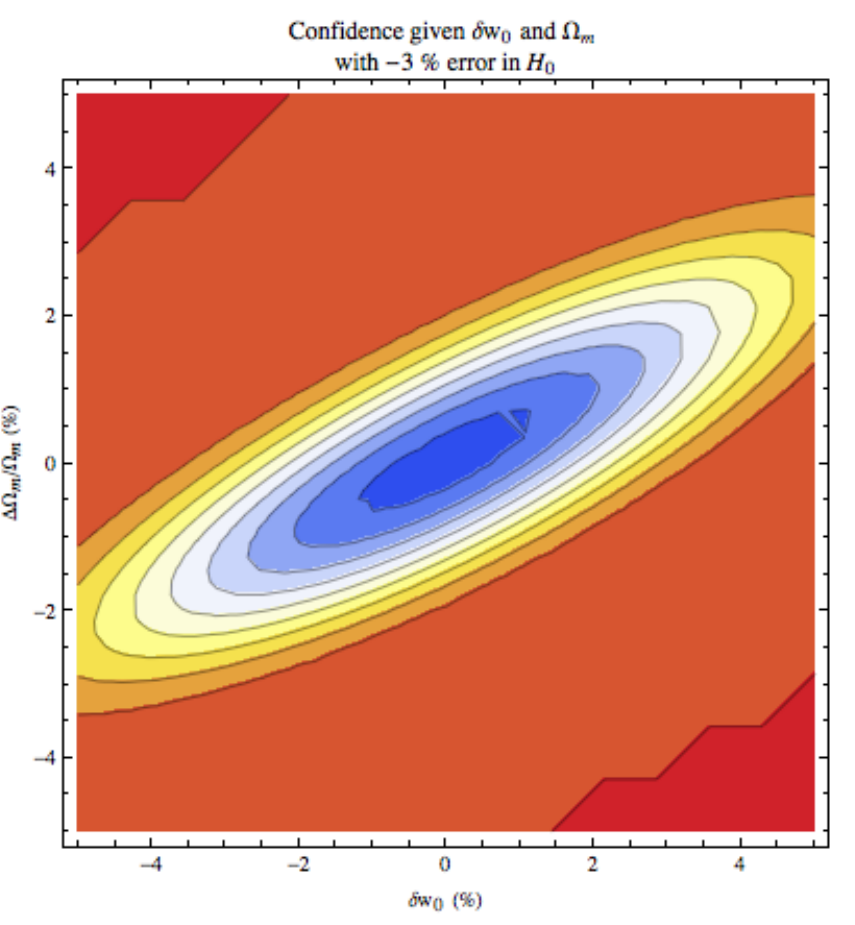} 
\centering{}\caption{The confidence of a detection as a function of $\delta w_0$ and $\Omega_m$ assuming a $-3\%$ error in $H_0$. Warmer colors denote a higher-confidence detection of slow-roll DE; the scale should be read just like a thermometer. The orange ellipse intersecting $(-5,-3.5)$ is $95\%$ confidence; the white ellipse $75\%$, and the center blue $55\%$.  Note that this plot is very similar to that for an over-estimate  for $H_0$ of $3\%$; this illustrates that the $\chi^2$ depends mainly on the magnitude of the error and not the sign. However, the scaling eqn. (\ref{e:chisqvsh}) of \S10.4 shows that there is also a small contribution to the $\chi^2$ from a term linear in $\Delta H_0/H_0$; this term will allow the sign of the error in $H_0$ to affect the results.  It is this that underlies the slight differences between this Figure and Figure 9, which may especially be seen around $\Delta \Omega_m/\Omega_m\simeq 4\%$ and $\delta w_0 \simeq -4\%$.}
 \vspace{0.5in}
\label{f:res3}
\end{figure}

\begin{figure}       
\includegraphics[scale=.70]{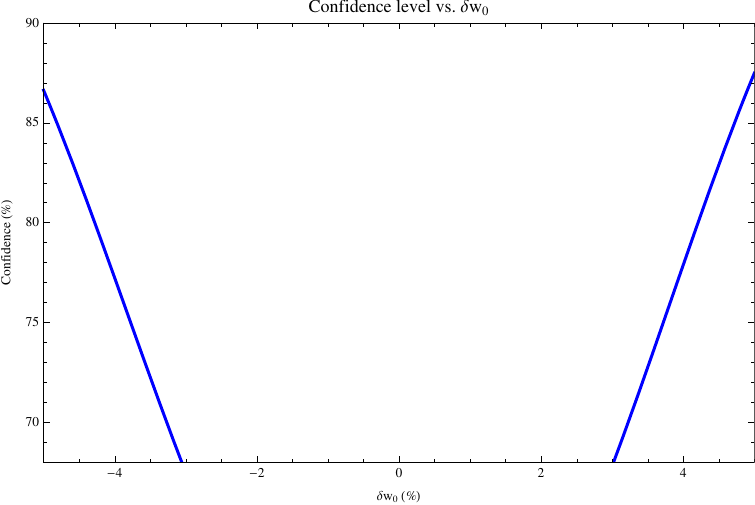} 
\centering{}\caption{The confidence level to which a detection of slow-roll DE is possible accounting for the possibility of errors in $\Omega_m$ and $H_0$.}
 \vspace{0.5in}
\label{f:res4}
\end{figure}

Figures 8-10 show the confidence values as a function of the percent change in matter density and the percent change in $\delta w_0$, the latter being equal to $\delta w_0$ because $\delta w_0$ is already normalized to $w=-1$.  There are two key points here.  First, one can view the ratio of the axes of the ellipse as an effective measure of how strongly each parameter ($\Omega_m$ and $\delta w_0$) comes into the final confidence estimate.  We focus now quantitatively on Figure 8, though our comments will apply qualitatively to Figures 9 and 10 as well.  The ratio of semi-major to semi-minor axis is $\simeq 2.9$, meaning roughly  that a given fractional change in $\Omega_m$ will have about $3$ times the effect on the confidence value that the same fractional change in $\delta w_0$ would have.  This is unfortunate because it essentially means the observations are more sensitive to the value of $\Omega_m$ than to the DE equation of state!  More concretely, this can be conceived as follows.  Suppose one is certain about $\Omega_m$ and believes one has detected $\delta w_0=2\%$ with confidence such that one is in the red region of the plot.  Now suppose one realizes there in fact $is$ some uncertainty about $\Omega_m$. Since the ellipse is tipped, it does not take much movement in $\Omega_m$ to move from the red region of the plot, where one has a high-confidence detection, to the blue region of the plot, where one has much less confidence. 


Second, the ellipse's major axis is offset by an angle of $\simeq \pi/7$ radians from the $\delta w_0$ axis.  This is a result of coupling between $\Omega_m$ and $\delta w_0$, a coupling introduced because we have constrained $\Omega_{m}+\Omega_{DE}=1$, so $\Omega_{DE}=\Omega_{m}-1$.  This coupling contributes to the fact that there exist particular erroneous values of $\Omega_m$ that can mimic the effect of a given $\delta w_0$.  The physical reason for this latter point, as we have already noted, is that matter has equation of state $p/\rho=0$, so adding extra matter pulls the average equation of state in a cosmological constant model up towards zero from $-1$.  Slow-roll DE with $\delta w_0>0$ has equation of state $p/\rho>-1$, so it may be mirrored by adding extra matter in a cosmological constant cosmology.  Phantom DE with $\delta w_0<0$, which has an equation of state $p/\rho<-1$, may analogously be mirrored by an underestimate of the matter density, letting the average total equation of state in a cosmological constant model tend closer to $-1$.

The comments above may be used to derive a formula for $\Omega_{mm}=\Omega_m+\Delta \Omega_m$ as a function of $\delta w_0$.  This formula is 
\begin{equation}
\Delta\Omega_m/\Omega_m\simeq\frac{\delta w_0}{\left(a/c\right)\cos(\theta)},
\label{e:ellipseeqngood}
\end{equation}
where $a$ and $c$ are respectively the semi-major and semi-minor axes of a representative ellipse in Figure 8 (chosen based on the confidence level of detection required) and $\theta$ is the angle between the horizontal axis of the plot and the ellipse's semi-major axis.
Using this to predict $\Omega_{mm}$ for $\delta w_0=3.5\%$ yields $\Omega_{mm}$ $1.34\%$ larger than the true matter density $\Omega_m=.272$.  The numerical value from \S5 is that $\Omega_{mm}$is $2.55\%$ larger than the true matter density. (Both numbers are ignoring errors in $H_0$ because they come from considering Figure 8, which is at constant $H_0$ equal to the current best value.)

Eqn. (\ref{e:ellipseeqngood}) encodes two effects. First, as we have already discussed, the confidence is more sensitive to the matter density than to the DE equation of state, so one needs to move less in matter to mirror a given move in equation of state; this is the factor of $(a/c)$.  Second, because the matter and DE densities are coupled by the flatness constraint, the DE equation of state will be coupled to the matter density (also as noted above).  Thus moving along the matter density-error axis is not the most efficient way of moving from one confidence contour to another on the ellipse: i.e. changing only the error in the matter density does not change the confidence of a detection as much as it would were there no coupling.  The most efficient route would be along the semi-minor axis of the ellipse, which is a gradient of the contour plot as is evident because it is perpendicular to the contours.  Because the axes of the plot do not align with this most efficient route, one pays a penalty of $\cos(\theta)$: for a given change in matter density $\Delta \Omega_m$, the $\delta w_0$ this mimics will be suppressed by a factor of $\cos(\theta)$. 

 One may fit ellipses to the confidence contours in Figure 8 analytically (see \S10.4 for details). The ellipses are given by
\begin{equation}
.428\delta w_0^2-1.052\delta w_0 \big( \Delta \Omega_m /\Omega_m \big)+1.319 \big( \Delta \Omega_m/\Omega_m \big)^2=-F
\label{e:ellipsereconst}
\end{equation}
where $-F>0$. All ellipses in the Figure are given by eqn. (\ref{e:ellipsereconst}) but with different values of $F$.  

Figure 11 takes the minimum confidence value associated with each $\delta w_0$ as the confidence with which a detection of that value might be claimed (see \S5.3.7 for details of this minimization).  Figure 11 shows that if $\delta w_0 =3.5\%$, a detection at $73\%$ confidence should be possible with upcoming experiments even if we are wrong about the matter density and $H_0$ in the least favorable way for detecting slow-roll DE.  If $\delta w_0 =5\%$, a possibility still very much observationally allowed, then the confidence of a detection even in this worst-case scenario would rise to $\simeq 85\%$.


\section{Conclusion}
\label{sec:conclusion}

In this paper, we have made two major arguments.  First, we have suggested that, if DE is {\it not} a cosmological constant with equation of state $w\equiv -1$, the simplest alternative would be that it is a second epoch of inflation, driven by a mechanism similar to that likely behind the first---a scalar field slowly rolling down the hill of its potential (i.e. in slow-roll).  Should Planck detect the tensor mode amplitude predicted by e.g. Linde's chaotic inflation, that would be smoking-gun evidence for a first epoch of inflation.  If that occurs, the prospect that DE may be a second epoch of inflation driven by an analogous mechanism should be taken seriously.  

We have here developed this idea to show that in such a DE model, the Hubble constant will have a generic evolution with redshift that is relatively insensitive to the starting value of the field's potential or the shape of the potential, and dependent solely on the difference from $-1$ in the DE equation of state today.  This differs from previous work in that previous work (e.g. Chiba 2009, Novosyadlyj 2010, Crittenden 2007) derived two-parameter forms for $w$.  By showing that $w$ is insensitive to the scalar field's acceleration as long as the acceleration is either small or roughly constant, in this work we have obtained a one-parameter model for $w$. 

We have used this result to assess whether observations upcoming in the next decade will be able to distinguish between slow-roll DE and cosmological constant, $w\equiv -1$, DE.  The current error bars from WMAP-7+BAO+$H_0$ constrain $w_0$ to be near $-1$ at roughly the 10\% level; we have been even more conservative in our estimates and considered deviations from $w=-1$ at present of $\lesssim 5\%$.  We find that, neglecting errors in the cosmological parameters $H_0$ and $\Omega_m$, if DE is a field in slow-roll with $w+1=3.5\%$ today, this would be detectable to $96\%$ confidence by observations in the next decade.  Accounting for the current error bars on the cosmological parameters $H_0$ and $\Omega_m$, this picture worsens somewhat.  For reasons we present in \S5, we find that, assuming a flat cosmology (motivated by inflation), the confidence level of a detection of $w$ different from $-1$ is only affected by errors in the matter density and $H_0$ today.  Detailed numerical modeling of the effects of such errors shows that a difference from $w=-1$ of $3.5\%$ could be detected with $73\%$ confidence (see Figure 11).  We have quantified the error introduced in our form for $H(z)$ (eqn. 14) due to the approximation new to this work, and shown that (Figure 19) for a $\phi^2$ potential the error will be an order of magnitude less than the signal.  Since our analytical form eqn. (51) for this error bound is generic to other potentials, and to order of magnitude matches the numerical results for $V\propto \phi^2$ (see Figure 18), the error introduced by our approximation should also be negligible compared to the signal for other potentials.

There is a second, parallel main thread to this work as well.  As noted in the Introduction, there are numerous time-varying physical models of DE.  Because testing each model individually would be overwhelming, parametrizations have been proposed that give the redshift-evolution of $w$ once their parameters have been fixed.  Each physical model may then produce predictions for the parameters in these parametrizations.  However, such parametrizations often build in a shape for the redshift-evolution of $w$ that has little physical motivation.  Therefore a formula that is physically motivated but still able to test a broad class of models is desirable.  We suggest that eqn. (14) of this work satisfies these desiderata.  As we have shown, it will apply to quintessence models of DE (these are the most analogous to the mechanism of inflation), and also to phantom DE, as long as in each model the field is, as expected, in slow-roll.  We already know from observation that one of the two conditions required for slow-roll is satisfied today:  since in all of these models, $w+1$ is proportional to the square of the field's velocity, the tight constraints on $w$'s deviation from $-1$ demand the field's velocity today be small. 

We have shown that, for both quintessence and phantom DE in slow-roll, eqn. (14) will describe the evolution of the Hubble constant with redshift.  Thus our calculations of the confidence with which slow-roll DE may be distinguished from a cosmological constant have broad implications: if DE is in slow-roll, we may expect an observational detection with some confidence in the next decade, dependent on the value of $w+1$ today.  

With this formula in hand, observers may be able to extract a signal from otherwise too-noisy data: this is because the shape of $w(z)$ in our formula is physically motivated: therefore whether the observations fit this shape provides additional information over and above the absolute amplitude of the curve, which is just set by $\delta w_0=w+1$ today.  Because our formula has only one free parameter, $\delta w_0$, it is more easily tested against observations than the two parameter models typically favored (e.g. the CPL parametrization) because it will be quickly evident if the prediction for $w(z)$ in the past implied by $\delta w_0$ today is not fulfilled.  Furthermore, with the shape for $w(z)$ our formula gives, data on $w(z)$ in the past may be used to tighten the constraints on $w$ today.  This is because this data, if fitted with the shape our formula implies, will demand a unique value of $w$ today. 

This work also points out that improvements in independent measurements of $H_0$ and $\Omega_m$ are very important to the study of DE.  Original methods such as the original LISA proposal for measuring $H_0$ (before its budget was cut) and measures of $\Omega_m$ from peculiar velocities and gravitational lensing, to name just a few, should be encouraged.  They provide important support work for the exciting DE observational programmes now proposed.

\section{Acknowledgements}

ZS thanks Paul Steinhardt, Daniel Eisenstein, Neta Bahcall, Harvey Tananbaum, Stephen Portillo, Philip Mocz, Aisha Down, and Laura Kulowski for many helpful conversations during the course of this work.  ZS particularly appreciates the willingness of the first on this list to be both critic and sounding board.  All authors thank Michael Blanton for helpful comments during revision. This material is based upon work supported by the National Science Foundation Graduate Research Fellowship under Grant No. DGE-1144152.  

\section{References}


\hspace{0.2in}Alam U, Sahni V and Starobinsky A, 2007, JCAP 0702:011

Albrecht A and Steinhardt P, 1982, Phys. Rev. Lett. 48 (17): 1220-1223

Albrecht A. et al.,  2006, Report of the
Dark Energy Task Force, arXiv:astro-ph/0609591

ALPACA website: \\
http://www.astro.ubc.ca/LMT/alpaca/index.html

Armendariz-Picon C, Mukhanov V and  Steinhardt P, 2001, Phys. Rev. D 63, 103510 

Blake C et al., 2005, Dark Energy Experiments with the Square Kilometer Array whitepaper,
http://www.astro.cornell.edu/~cordes/SKA/SKA\_DETF\_
whitepaper.pdf 

Cahn RN, de Putter R, and Linder EV, 2008, JCAP 0811:015

Caldwell RR, 2002, Phys. Lett. B545: 23-29

Caldwell RR, Kamionkowski  M and Weinberg NN, 2003, Phys. Rev. Lett. 91 071301

Caldwell RR and Linder EV, 2005, Phys. Rev. Lett. 95, 141301

Cardelli JA, Clayton GC and Mathis JS, 1989, ApJ 345: 245-256

Causse MB, 2004, arXiv:astro-ph/0312206


CFHLS website: \\
http://www.cfht.hawaii.edu/Science/CFHLS/

Chandrasekhar S, 1931, ApJ 74

Chevallier M and Polarski D, 2001, Int. J. Mod. Phys. D 10, 213 

Chotard et al., 2011, arXiv:1103.5300v1

Chiba T, 2009, Phys. Rev. D79, 083517

Chiba T, Dutta S and Scherrer R, 2009, Phys. Rev. D80: 043517 

Chiba T, Siino M and Yamaguchi M, 2010, Phys. Rev. D81: 083530

Chiba T, De Felice A and Tsujikawa S, 2013, Phys. Rev. D87: 083505

Chiba T, Okabe T and Yamaguchi M, 2000, Phys. Rev. D62: 023511


Cline J,  Jeon S and Moore G, 2004, Phys. Rev. D 70, 043543 

Copeland E, Sami M and Tsujikawa S, 2006, Int. J. Mod. Phys. D 15:1753-1936

Colombo LPL, Pierpaoli E and Pritchard JR, 2009, MNRAS 398:1621

Crittenden R, Majerotto E and Piazza F, 2007, Phys. Rev. Lett. 98:251301

Day C, 2010, Phys. Today 63(3), 33

De Boni C, Dolag K, Ettori S, Moscardini L, Pettorino V, Baccigalupi C, 2011, MNRAS, 415, 2758 

de Putter R, Zahn O, Linder E, 2009, arXiv:0901.0916

DES website:\\
http://www.darkenergysurvey.org/




Dutta S and Scherrer R, 2008, Phys. Rev. D78: 123525

Eisenstein DJ, 2013, ``Simple Forms for the Distance-Redshift Relation in Dark Energy Cosmologies'', private communication.


Euclid website:\\
http://sci.esa.int/science-e/www/object/index.cfm?\\
fobjectid=42266

Eisenstein DJ et al., 2005, ApJ 633: 560574

Eisenstein DJ , Seo H-J and White M, 2007, ApJ 664: 660-674

Eisenstein DJ et al., 2011, http://arxiv.org/abs/1101.1529



Geisbusch J and Hobson P, MNRAS (2007) 382 (1): 158-176.


Green J et al., 2012, Wide-Field InfraRed Survey Telescope WFIRST Final Report, http://wfirst.gsfc.nasa.gov/science/

Guth AH, 1981, Phys. Rev. D, 23, 2, 347-356 

Guth AH, 2007, J. Phys. A40: 6811-6826

Guth AH and Kaiser D, 2005, Sci. 307, 884-90

Gott JR et al., 2009, Astrophys. J., 695, L45

Gott JR and Slepian Z, 2011, MNRAS 416, 2, 907-916

Heavens A, 2009, Nucl. Phys. Proc. Suppl. 194 76

HETDEX website:\\ 
http://hetdex.org/


Hubble website:\\ 
http://hubblesite.org/hubble\_discoveries/

Ivezic Z et al., 2011, arXiv:0805.2366,
http://lsst.org/lsst/overview/ 

Kim A, 2004, LBNL Report LBNL-56164


Komatsu E et al., 2011, Astrophys. J. Suppl. 192:18

LAMOST website:\\
http://www.lamost.org

Lesgourgues J, 2006, Inflationary Cosmology, available at: 
lesgourg.web.cern.ch/lesgourg/Inflation\_EPFL.pdf

Li M, Li X, and Zhang X, 2011, Sci. China Phys. Mech. Astron. 53:1631-1645

Linde AD, 1982, Phys. Lett. B 108 (6): 389-393

Linde AD, 2002, report number SU-ITP-02-25, arXiv: hep-th/0205259

Linder E, 2003, Phys. Rev. Lett. 90, 091301 

Lyth D and Liddle A, 2000, Cosmological inflation and large-scale structure. Cambridge: University Press

Melchiorri A, Mersini L, Odman CJ and Trodden M, 2003, Phys. Rev. D 68, 043509

Neupane IP and Scherer C, 2008, JCAP 5 009

Noh Y, White M and Padmanabhan N, 2009, arXiv:0909.1802

Novosyadlyj B, Sergijenko O and Apunevych S, 2010, arXiv:1008.1943v2

Novosyadlyj B, Sergijenko O and Apunevych S, 2011,  Journal of Physical Studies, v. 15, No 1, id. 1901

Novosyadlyj B, Sergijenko O, Durrer R and Pelykh  V, 2012, Phys. Rev. D  vol. 86, Issue 8, id. 083008 

Novosyadlyj B, Sergijenko O, Durrer R and Pelykh  V, 2013, JCAP 06 id. 042 

Padmanabhan N and White M, 2009, Phys. Rev. D 80: 063508

Pan-STARRS website:\\ 
http://pan-starrs.ifa.hawaii.edu/public/

Park C and Kim Y-R, ApJL vol. 715, 2, L185 (2010)

Park C, Choi Y-Y, Kim J, Gott JR, Kim SS and Kim K-S, 2012, arXiv:1209.5659


Perlmutter S, 2011, ``Dark Energy: One Observer's Report Card'', talk at Princeton University

Perlmutter S, Turner MS and White M, 1999, Phys. Rev. Lett., 83, 670 

Perlmutter S et al. (The Supernova Cosmology Project), 1998, Nature, 391, 51

Phillips MM, 1993, ApJ 413 L105

Phinney ES, 2002, Science Requirements for LISA, http://www.its.caltech.edu/?esp/lisa/LISTwg1.req-pr.pdf

Planck: the scientific programme, 2005, available at:
http://www.rssd.esa.int/Planck




Refregier A, Amara A, Kitching TD, Rassat A, Scaramella R and Weller J, 2010, http://arxiv.org/abs/1001.0061

Riess A et al., 1998, AJ, 116: 1009-1038 

Riess A et al., 2011, ApJ, 730, 119

Sahni V, Shafieloo A and Starobinsky A, 2008, Phys. Rev. D 78:103502
 
SDSS-III: Massive Spectroscopic Surveys of the Distant Universe, the Milky Way Galaxy, and Extra-Solar Planetary Systems, 2008, http://www.sdss3.org/science.php

Seo H-J and Eisenstein DJ, 2007, ApJ, 665: 14-24

Slepian Z, 2011, Princeton University Thesis in Astrophysical Sciences.

Speare R, 2012, Princeton University Thesis in Astrophysical Sciences.

SPT  website:\\ 
http://pole.uchicago.edu/ 

Wang PY, Chen CW and Chen P, 2012, JCAP 1202, 016





Vanderlinde K et al., 2010, ApJ 722 1180-1196

Vilenkin A, 2003, Int. J. Theor. Phys. 42, 1193-1209

Weinberg DH, Mortonson MJ, Eisenstein DJ, Hirata C, Riess AG, and Rozo E, 2013, Physics Reports, vol. 530, Issue 2, 87-255 

Weinberg S, 1987, Phys. Rev. Lett. 59, 2607-2610


\section{Appendix A}
\label{sec:appendix}

\subsection{Self-consistency of the approximations of \S2, and numerical results}
In a single component universe with only DE, the approximation that $H$ evolves more in time than $\partial V/ \partial  \phi$ would not be correct, as $H\propto\sqrt{\rho}\approx \sqrt{V}$.  For instance, for a quadratic potential, $H\propto \phi$ and $\partial V/ \partial  \phi\propto \phi$ as well.  Similarly, in inflationary cosmology, this approximation would be incorrect because the inflaton field is the sole non-negligible source of the energy density.  However, for $z\sim 0-2$, {\it matter} also determines the time-evolution of $H$.  Thus $H$ may evolve in time even if $V$ and $\partial V/\partial \phi$ do not, and the approximation is self-consistent.  

To show this quantitatively, we ask which term makes the dominant contribution to the time-evolution of $\dot{\phi}$: the Hubble constant or the slope of the potential?  We have for a standard scalar field

\begin{eqnarray}
\frac{d\dot{\phi}}{dz} = -\frac{1+SR2}{3H} \frac{d}{d\phi} \bigg\{ V\bigg(\frac{dV/dz}{V}-\frac{H_0^2}{2H^2}\bigg[3\Omega_m(1+z)^2+ \nonumber \\
\frac{3\Omega_{DE} \delta w}{1+z} \exp \left[ 3\int_0^z \frac{\delta w}{1+z'} dz' \right] \bigg]  \bigg) \bigg\},
\label{e:cons1}
\end{eqnarray}
from rearranging and differentiating the relation directly after eqn. (4).  Note that here we have taken $SR2\equiv \ddot{\phi}/V'$ to be independent of $z$, which is valid, as Figs. 13 and 15 show, for the regime in which $SR2$ is not negligible compared to unity.  As these Figures show, $SR2$ does indeed vary with $z$ for $z \lesssim 1$. However, in this latter case $SR2$ is roughly negligible compared to unity, so it drops out of eqn. (\ref{e:cons1}).  Finally, we have also used eqn. (10) to find $dH/dz$ and simplified.  

We wish to show that for small $\delta w\ll1$, the first term  in the parentheses within the curly brackets is much less than the second term, as the first term represents the contribution of $\partial V/ \partial  \phi$ to the evolution of $\dot {\phi} $ and the second term represents the contribution of $H$.  For the first term, we have

\begin{eqnarray}
\left|\frac {dV/dz}{V}\right| =\left|\frac{1}{V}\frac{\partial V}{\partial  \phi} \dot \phi \frac{a}{H} \right|=\frac{3a\delta w}{1+SR2} \lesssim 0.225.
\label{e:firstterm}
\end{eqnarray}

The first equality uses the chain rule, while the second uses the relation directly after eqn. (4) to replace $\partial V/\partial \phi$ with $-3H\dot{\phi}/\left(1+SR2\right)$ and then that $\delta w\approx \dot\phi^2/V$. The final bound uses $a\leq1$, $\delta w\leq\delta w_0\leq .05$, and ${\rm min}\;SR2=-1/3$, the last from Figs. 13 and 15 . Note that in the limit that $SR2\ll 1$, which holds as $z\rightarrow 0$, the bound becomes sharper: 0.225 becomes 0.15.

For the second term in the curly brackets in eqn. (\ref{e:cons1}), we want a minimum, because we wish to show that the second term will dominate the first term for the redshifts we consider.  Setting $\delta w=0$ in the numerator (for a minimum), using that $\delta w$ is small in the denominator (given as $H=H(z)$) to expand, and dropping higher-order terms (in $\Omega_{DE}\delta w$), we find 
\begin{eqnarray}
& {\rm min}\bigg\{\frac{H_0^2}{2H^2}\bigg[3\Omega_m(1+z)^2+ \frac{3\Omega_{DE}\delta w}{1+z}\exp\left[3\int_0^z\frac{\delta w}{1+z'}dz'\right]\bigg]\bigg\} \nonumber \\
& \approx \frac{3\Omega_m(1+z)^2)}{2(\Omega_m(1+z)^3+\Omega_{DE})}\bigg|_{z=0}=0.4.
\label{e:secondterm}
\end{eqnarray} 
Note that we evaluate at $z=0$ because that is where the minimum of the expression above occurs.

Comparing eqns. (\ref{e:firstterm}) and (\ref{e:secondterm}) shows that for a field in slow-roll, our approximations in \S2 will hold. This analysis also applies to phantom DE. In those models, the equation of motion for the field, and $\delta w$, differ only in sign from standard scalar field models. Our analysis here is sensitive to the magnitudes of quantities but not their signs.  

We now present several figures (Figures 12-15) illustrating how well the approximate formulae eqns. (14) and (15) work for typical potentials in each model.  Here our goal is not to provide comprehensive coverage of every possible initial condition or potential, but rather to offer examples to supplement and bolster the  analytical work we have already discussed.  We briefly detail the numerical method used to obtain these figures later in this Appendix (\S10.6). 

We plot the slow-roll parameters  (defined following Lesgourgues 2006) ${\rm SR1} \equiv \frac{1}{2}\dot{\phi} ^2/V$ and ${\rm SR2} \equiv \ddot {\phi} / ( \partial V /\partial \phi)$ in Figures 13 and 15; these show the fields actually are in slow-roll for typical potentials in each model we simulate. Figures 12 and 14 show the applicability of eqn. (15) to typical models.

\begin{figure}       
\includegraphics[scale=.6]{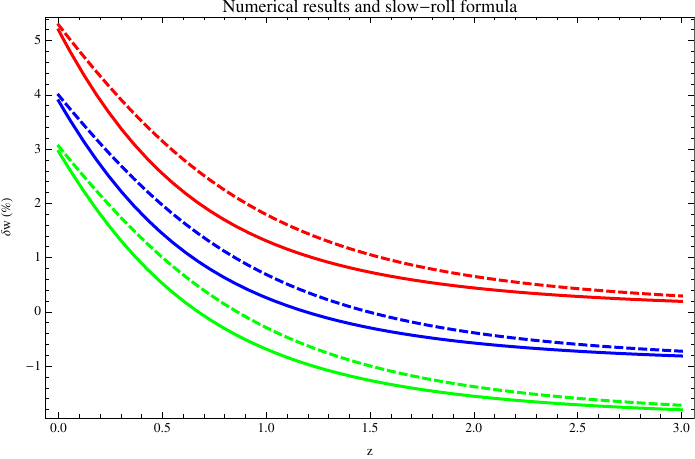} 
\centering{}\caption{Scalar field: The solid curves are the results of exact, self-consistent numerical solution of the Friedmann equation and scalar field equation of motion for several typical potentials.  Dashed is the slow-roll formula eqn. (9) evaluated using eqn. (14) for the Hubble constant (i.e., eqn. (15)). Red is for a quadratic potential, blue for a quartic potential, and green for an exponential potential. All curves correspond to $\delta w_0\approx 5\%$; we have shifted the quartic result down by a constant $1\%$ and the exponential result down by a constant $2\%$ so that all three curves are clearly visible. The quadratic test was done in Gott \& Slepian 2011.}
 \vspace{0.5in}
\label{f:scalarfieldresults}
\end{figure}

\begin{figure}       
\includegraphics[scale=.6]{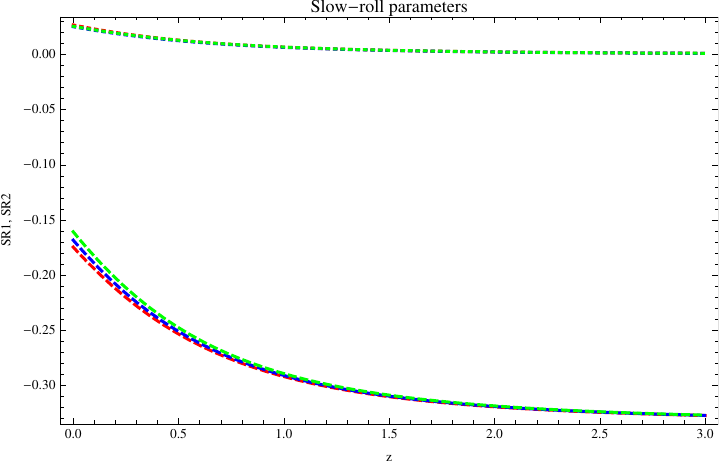} 
\centering{}\caption{Scalar field: These show the 2 slow-roll parameters for each potential.  Red is quadratic, blue is quartic, green is exponential.  Dotted is SR1 (all greater than zero), dashed is SR2.  All three curves overlap for SR1 so that they are indistinguishable; they also overlap for SR2 but can be distinguished.The key point is that all of the curves have magnitude much less than unity, meaning the fields are in slow-roll.}
 \vspace{0.5in}
\label{f:scalarfieldslowroll}
\end{figure}

\begin{figure}       
\includegraphics[scale=.6]{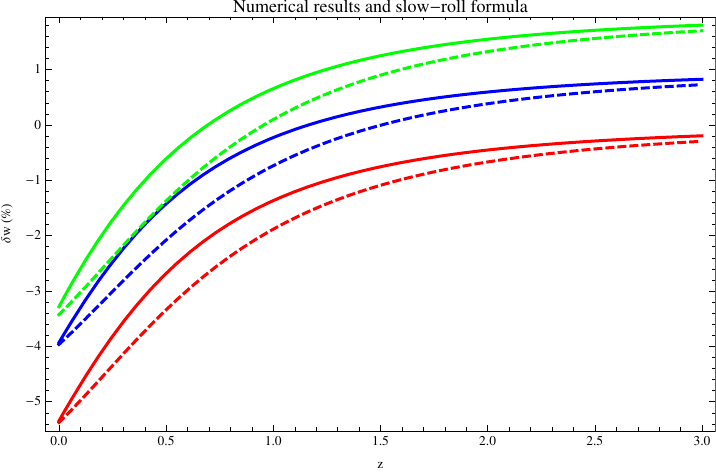} 
\centering{}\caption{Phantom: The solid curves are the results of exact, self-consistent numerical solution of the Friedmann equation and phantom field equation of motion for several typical potentials.  Dashed is the slow-roll formula eqn. (9) evaluated using eqn. (14) for the Hubble constant (i.e., eqn. (15)). Red is for a quadratic potential, blue for a quartic potential, and green for an exponential potential. All curves correspond to $\delta w_0\approx 5\%$; we have shifted the quartic result up by a constant $1\%$ and the exponential result up by a constant $2\%$ so that all three curves are clearly visible.}
 \vspace{0.5in}
\label{f:phantomresults}
\end{figure}

\begin{figure}       
\includegraphics[scale=.7]{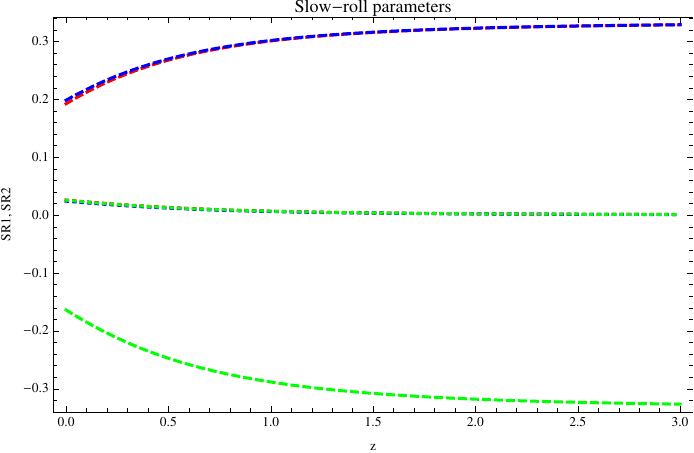} 
\centering{}\caption{Phantom: These show the 2 slow-roll parameters for each potential.  Red is quadratic, blue is quartic, green is exponential.  Dotted is SR1 (these curves begin with values $\simeq 0.02$ for $z=0$), dashed is SR2.   All three curves overlap for SR1 so that they are indistinguishable; two of the three also overlap for SR2 but can be distinguished. The key point is that all of the curves have magnitude much less than unity, meaning the fields are in slow-roll.}
 \vspace{0.5in}
\label{f:phantomslowroll}
\end{figure}

\subsection{Validity of the approximations of \S2}

Recall the exact equation of motion eqn. (4) may be written

\begin{equation}
3H\dot{\phi}=-\frac{\partial V}{\partial\phi}\left[1+SR2\right],\end{equation}

with $SR2\equiv\ddot{\phi}/\left(\partial V/\partial\phi\right)$.
There are two regimes: 

\textbf{1) }when $SR2$ varies noticeably with redshift, which occurs
for $z<1$, and

\textbf{2) }when $SR2$ is nearly constant and approximately equal
to $-1/3,$ which occurs for $z\gtrsim1$.

In each regime, to verify our scaling $\dot{\phi}\propto1/H,$
we must check two conditions: 

\textbf{i.} that $SR2$'s explicit presence above does not invalidate
the scaling, and 

\textbf{ii. }that $SR2$ does not cause $\partial V/\partial\phi$
to vary comparably to $1/H$. 

We have already checked \textbf{ii.} for both regime \textbf{1)} and regime \textbf{2)} in \S10.1, so we here focus on \textbf{i.}

For regime \textbf{1)}, where $SR2$ varies noticeably with redshift,
we approximate $SR2\approx sz+s_{0},$ where $|s|\lesssim.15.$  Now, we have

\begin{equation}
\dot{\phi}=-\frac{\partial V/\partial\phi\left[1+s_{0}\right]}{3H}-\frac{\partial V}{\partial\phi}\frac{sz}{3H}.
\label{e:102e1}
\end{equation}

The first term will just yield $\dot{\phi}\propto1/H.$ Since we fix
the normalization of this scaling by setting $\delta w(z=0)=\delta w_{0},$
our original formula eqn. (9) will capture this first term's behavior
with no error. We now calculate the correction to $H(z)$ from the
second term in eqn. (\ref{e:102e1}). We have

\begin{equation}
\delta w\approx\frac{\dot{\phi}^{2}}{V}\approx\left(-\frac{\partial V/\partial\phi}{3}\right)^{2}\left[\frac{\left(1+s_{0}\right)^{2}}{H^{2}}+\frac{2(1+s_{0})sz}{H^{2}}+\frac{\left(sz\right)^{2}}{H^{2}}\right].
\label{e:102e2}
\end{equation}

For the second and third terms in square brackets in eqn. (\ref{e:102e2}), we neglect
terms of order $\left(sz\right)^{2}$ and $s_{0}sz,$ so we find

\begin{equation}
\delta w\approx C\left[\frac{\left(1+s_{0}\right)^{2}}{H^{2}}+\frac{2sz}{H^{2}}\right].\end{equation}

As already indicated, the first term will be perfectly described by
our original formula since the normalization of $\delta w$ is fixed
separately and the term in $z$ vanishes at $z=0$. We thus have

\begin{equation}
\delta w(z)\approx\frac{\delta w_{0}H_{0}^{2}}{H^{2}(z)}+\frac{2H_{0}^{2}\delta w_{0}}{\left(1+s_{0}\right)^{2}}\frac{sz}{H^{2}(z)}=\frac{\delta w_{0}H_{0}^{2}}{H^{2}(z)}\left[1+\frac{2sz}{\left(1+s_{0}\right)^{2}}\right],\end{equation}

where we have used that $\delta w_{0}=C\left[\left(1+s_{0}\right)^{2}/H_{0}^{2}\right]$
to fix $C=H_{0}^{2}\delta w_{0}/\left(1+s_{0}\right)^{2}$ today and
that $\partial V/\partial\phi$ is roughly constant compared to $1/H$ (cf. \S10.1) to see that $C$ is roughly constant. We now wish to see what correction
the linear term in $z$ in the square brackets above will produce to eqn. (14). 

We calculate the
correction to the argument of the exponential in eqn. (10) as 

\begin{equation}
r(z)\equiv\frac{3}{\left(1+s_{0}\right)^{2}}\int_{0}^{z}\frac{dz'}{1+z'}\frac{2sz'}{\Omega_{m}(1+z')^{3}+\Omega_{DE}}.\end{equation}

Evaluating the integral,

\begin{eqnarray}
r(z)=\frac{s}{\Omega_{DE}\Omega_{m}^{1/3}\left(1+s_{0}\right)^{2}}\bigg \{ \ln\left[\alpha^{-\Omega_{DE}^{1/3}}\beta^{2\Omega_{DE}^{1/3}}\gamma^{2\Omega_{m}^{1/3}}\right] \nonumber \\
+2\sqrt{3}\Omega_{DE}^{1/3}\tan^{-1}\delta-\zeta\bigg\} \bigg|_{0}^{z},
\end{eqnarray}

with

\begin{equation}
\alpha(z)\equiv\Omega_{DE}^{2/3}-\left(\Omega_{DE}\Omega_{m}\right)^{1/3}\left(1+z\right)+\Omega_{m}^{2/3}\left(1+z\right)^{2},\end{equation}

\begin{equation}
\beta(z)\equiv\Omega_{DE}^{1/3}+\Omega_{m}^{1/3}\left(1+z\right),\end{equation}

\begin{equation}
\gamma(z)\equiv\Omega_{DE}+\Omega_{m}\left(1+z\right)^{3},\end{equation}

\begin{equation}
\delta(z)\equiv\frac{2\Omega_{m}^{1/3}\left(1+z\right)-\Omega_{DE}^{1/3}}{\sqrt{3}\Omega_{DE}^{1/3}},\end{equation}

and \begin{equation}
\zeta(z)\equiv6\Omega_{m}^{1/3}\ln\left(1+z\right).\end{equation}

We may now write the dark energy term in eqn. (14) for $H^{2}$ as 

\begin{equation}
\Omega_{DE}e^{\delta w_{0}g\left(1+\epsilon\right)}\approx\Omega_{DE}e^{\delta w_{0}g}\left(1+\delta w_{0}r\right),\end{equation}

where we have defined 

\begin{equation}
g(z)=\frac{1}{\Omega_{DE}}\ln\left[\frac{\left(1+z\right)^{3}}{\Omega_{m}\left(1+z\right)^{3}+\Omega_{DE}}\right]\end{equation}

and \begin{equation}
\epsilon(z)=r(z)/g(z)\ll1,\end{equation}

so the approximate equality above follows from Taylor expanding the
exponential about $\epsilon=0.$ The term proportional to unity in
the corrected dark energy term (eqn. (46)) will give the original result eqn.
(14), so we have

\begin{equation}
\Delta H_{SR2,lin}^{2}\approx\frac{H_{0}^{2}\Omega_{DE}e^{\delta w_{0}g(z)}\delta w_{0}r(z)}{H^{2}(z)}\end{equation}

where $\Delta H_{SR2,lin}^{2}$ is just the fractional correction
to $H^{2}$ due to $SR2$ (in the regime where $SR2$ is the linear
function of $z$ $SR2\approx s_{0}+sz$) and $H^{2}$ is given by
eqn. (14). We then have that

\begin{equation}
H_{corr,SR2,lin}=H\left(1+\Delta H_{SR2,lin}^{2}\right)^{1/2}\approx H\left(1+\frac{1}{2}\Delta H_{SR2,lin}^{2}\right)\end{equation}

so that the fractional difference between $H$ as given by eqn. (14)
and $H_{corr,SR2,lin}$ is

\begin{equation}
\delta H\approx\frac{1}{2}\Delta H_{SR2,lin}^{2}.\end{equation}

We plot this using $s=-.15$ and $s_{0}=-.15$ in Figure 16.

\begin{figure}

\includegraphics[scale=.5]{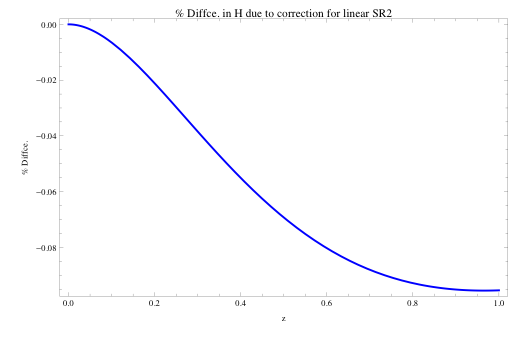}\caption{$\delta H$ (\%) as given by eqn. (51).}

\end{figure}

We now treat regime \textbf{2)},\textbf{ }where $SR2\approx{\rm const\approx-1/3}$.  Note that we have chosen
a specific normalization for the scaling $\delta w\propto 1/H^2$ by setting it to  $\delta w_0$ at $z=0.$
This incorporates the constant $s_{0}$ in our expansion for $SR2$
in regime \textbf{1)}, but here $SR2\approx{\rm const}=-1/3\neq s_{0}$.
The fact that $SR2$ now becomes a constant not equal to $s_{0}$
means that $\delta w=\delta w_{0}H_{0}^{2}\chi/H^{2}(z)$, where $\chi$
accounts for the fact that the scaling now must have a different constant
of proportionality. We may easily obtain the error by letting $z=1.22$
in the numerator of $\Delta H_{SR2,lin}^{2}$ in our relation for
$\delta H$, because at this $z$ $SR2=1/3,$ and $SR2$ is a continuous
function at the boundary between the two regimes. $H^{2}(z)$ in the
denominator of $\Delta H_{SR2,lin}^{2}$ continues to increase with
$z,$ however. Thus we may extend our error plot out to higher $z.$ See Figures 17 and 18.  In Figure 19 we compare the error in $H$ due to our formula's being approximate to the signal we seek (the difference between slow-roll and cosmological constant cosmologies), showing that this signal is always approximately an order of magnitude larger.

\begin{figure}
\includegraphics[scale=.4]{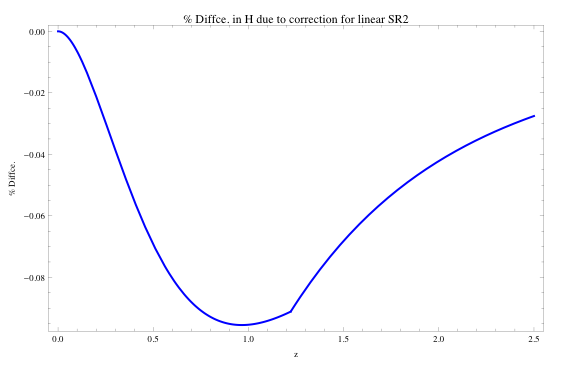}\caption{$\delta H$ (\%) as given by eqn. (51), extended to regime
\textbf{2) }as explained in the text.}

\end{figure}

\begin{figure}

\includegraphics[scale=.4]{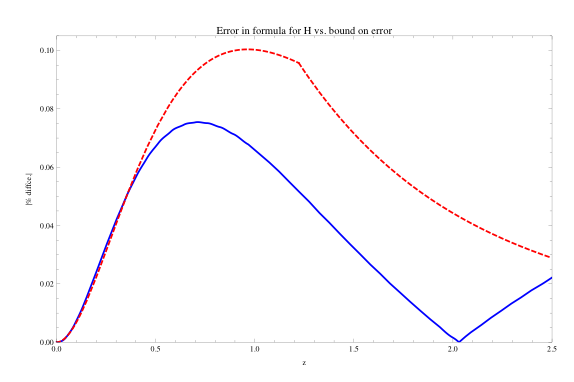}\caption{Here we plot the absolute values of the error between our formula
for $H$ eqn. (14) and the exact numerical results for $\delta w_{0}\simeq4.\%$
(blue, solid) and the bound on this error given by eqn. (51) and extended as explained in the text (red, dashed). The cusp in the numerical results is because the error goes negative for $z>2$ and we have taken an absolute value. Note the precise behavior of the error of our formula here should be unimportant because these points contribute little to the $\chi^2$, both because few observations go to $z>2$ and because DE becomes subdominant to matter for $z\gtrsim 2$.}

\end{figure}

\begin{figure}

\includegraphics[scale=.4]{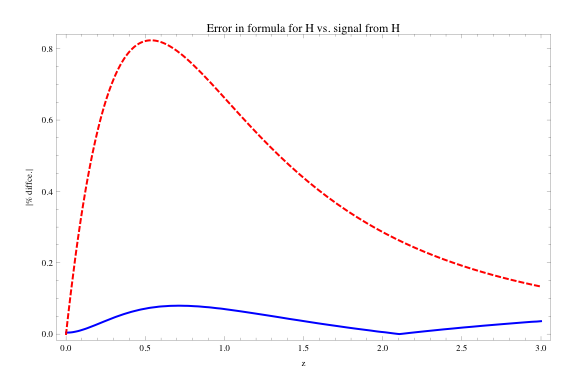}\caption{Here we plot the absolute value of the error in $H$ between our formula
eqn. (14) and the exact numerical results (blue, solid) as compared
to the difference between our slow-roll formula for $H$ and a cosmological
constant $H$, the signal we hope to detect (red, dashed). For $z\lesssim2,$
the error is always approximately an order of magnitude less than
the signal, meaning our formula eqn. (14) will be a fairly accurate
descriptor of how a slow-roll DE model differs from a cosmological
constant DE model.}

\end{figure}


\subsection{Fractional differences in Hubble constant and comoving distance}
\label{subsec:approxdadl}

We can use eqn. (\ref{e:slowrollfman}) to compute analytical expressions for the fractional difference in $d_L$ and $d_A$, the luminosity and angular diameter distances, from those for a $w\equiv -1$ cosmology. We define
\begin{equation}
\frac{\Delta d_L}{d_{L,-1}}\equiv\frac{d_{L,SR}-d_{L,-1}}{d_{L,-1}}
\label{e:defdeltadL}
\end{equation}
and
\begin{equation}
\frac{\Delta d_A}{d_{A,-1}}\equiv\frac{d_{A,SR}-d_{A,-1}}{d_{A,-1}},
\label{e:defdeltadA}
\end{equation}
where $d_{L,SR}$ is $d_L$ for a slow-roll DE cosmology and $d_{L,-1}$ is $d_L$ for a $w\equiv-1$ cosmology, and the same for $d_A$.
In a flat cosmology, $d_L$ and $d_A$ are both defined in terms of the comoving distance $d_c$ as follows:
\begin{equation}
d_L\equiv (1+z)d_c
\label{e:defdL}
\end{equation}
and
\begin{equation}
d_A\equiv(1+z)^{-1}d_c.
\label{e:defdA}
\end{equation}
See for example Copeland et al. 2006.
Evidently,
\begin{equation}
\frac{\Delta d_L}{d_{L,-1}}=\frac{d_{c,SR}-d_{c,-1}}{d_{c,-1}}=\frac{\Delta d_A}{d_{A,-1}}.
\label{e:comoveq}
\end{equation}
Therefore we seek $\Delta d_c/d_{c,-1}$, defined by the middle equality above.
Setting $a=1$ today and in units where $c=1$, the comoving distance is 
\begin{equation}
d_c\equiv\int_0^z\frac{dz'}{H}.
\label{e:comovdef}
\end{equation}
Working in the limit of small $z$, we can obtain an explicit formula for $\Delta d_c/d_{c,-1}$.  We have
\begin{eqnarray}
\Delta d_c\equiv d_{c,SR}-d_{c,-1}=\int_0^z\left[\frac{1}{H_{SR}}-\frac{1}{H_{-1}}\right]dz'\\
= - \int_0^z\frac{\Delta H/H_{-1}}{H_{SR}}dz'\nonumber,
\label{e:dcH}
\end{eqnarray}
where we have simplified and used the definition $\Delta H/H_{-1}=\left(H_{SR}-H_{-1}\right)/H_{-1}$ to obtain the second equality.
This expression is generally valid.  

Using the Friedmann equation with $w=-1$ and eqn. (\ref{e:slowrollfman}), both in the limit $z  \ll 1$ (and with $\Omega_{DE}+\Omega_m=1$), it is easily shown that
\begin{equation}
H_{SR}\approx H_0\left( 1+\frac{3}{2}z\left(\Omega_m+\Omega_{DE}\delta w_0\right)\right)
\end{equation}
and
\begin{equation}
H_{-1}\approx H_0\left(1+\frac{3}{2}\Omega_m z\right), 
\label{e:Hn1approx}
\end{equation}
so that 
\begin{equation}
\Delta H\equiv H_{SR}-H_{-1}\approx H_0\left(\frac{3}{2}\Omega_{DE}\delta w_0 z\right)
\label{e:Hdiffdef}
\end{equation}
and 
\begin{equation}
\Delta H/H_{-1} \approx \frac{3}{2}\Omega_{DE}\delta w_0 z\quad(small\;z).
\end{equation}
Using these results, it follows that 
\begin{equation}
\frac{\Delta H/H_{-1}}{H_{SR}}\approx\frac{3\Omega_{DE}\delta w_0 z}{2H_0}.
\end{equation}
Substituting this into eqn. (44) yields
\begin{equation}
d_{c,SR}-d_{c,-1} \approx -\frac{3\Omega_{DE}\delta w_0 z^2}{4H_0}.
\label{e:smallzdiffindc}
\end{equation}

We now seek $d_{c,-1}$, the denominator of $\Delta d_c/d_{c,-1}$. 
\begin{equation}
d_{c,-1}=\int_0^z\frac{dz'}{H_{-1}}\approx\frac{2}{3\Omega_m H_0}\ln\left[1+\frac{3}{2}\Omega_m z\right],
\end{equation}
where for the approximate equality we used eqn. (\ref{e:Hn1approx}) for $H_{-1}$ in the limit of  $z  \ll 1$.
Since we are in this limit, we further approximate $\ln(1+x)\approx x$, yielding
\begin{equation}
d_{c,-1}\approx \frac{z}{H_0}.
\label{e:dcn1}
\end{equation}

Combining  this result with eqn. (\ref{e:smallzdiffindc}), we find
\begin{equation}
\frac{\Delta d_c}{d_{c,-1}}\approx -\frac{3\Omega_{DE}\delta w_0}{4}z\quad(small\;z).
\label{e:smallzdeltadc}
\end{equation}
This result is shown in Slepian 2011.

\subsection{Scalings for $\chi^2$}

Taylor-expanding eqn. (\ref{e:slowrollfman}) about $\delta w_{0}=0$ yields 
\begin{equation}
H_{SR}\approx H_{-1}+\frac{H_{0}^{2}}{2H_{-1}}\ln\left[\frac{(1+z)^{3}}{\Omega_{m}(1+z)^{3}+\Omega_{DE}}\right]\delta w_{0}.
\label{eq:Taylordeltaw}
\end{equation}

This allows us to find the difference and fractional difference in
$H$ from that for a cosmological constant cosmology.

\begin{equation}
\Delta H\equiv H_{SR}-H_{-1}\approx\frac{H_{0}^{2}}{2H_{-1}}\ln\left[\frac{(1+z)^{3}}{\Omega_{m}(1+z)^{3}+\Omega_{DE}}\right]\delta w_{0}.\end{equation}

\begin{equation}
\frac{\Delta H}{H_{-1}}\approx\frac{1}{2}\left(\frac{H_{0}}{H_{-1}}\right)^{2}\ln\left[\frac{(1+z)^{3}}{\Omega_{m}(1+z)^{3}+\Omega_{DE}}\right]\delta w_{0}.
\label{e:dHoverHchi}
\end{equation}

We can obtain expressions for $\Delta d_{c}$ and $\Delta d_{c}/d_{c,-1}$ as well.  We approximate that $H_{SR}\approx H_{-1}$ in the denominator of the integrand of eqn. (44).  Using eqn. (\ref{e:dHoverHchi}) for the numerator of the integrand in eqn. (44) we find
\begin{equation}
\Delta d_{c}\approx\frac{\delta w_{0}}{6H_{0}}\left(I\left[(1+z)^{3}\right]-13\right),
\label{e:deltdcchi1}
\end{equation}

where \begin{eqnarray}
I[u]\equiv\bigg[u^{1/3}\bigg\{ \bigg[ { _2F_{1}}\bigg(\frac{1}{3},\frac{5}{6};\frac{4}{3};\frac{\Omega_{m}u}{\Omega_{m}u +\Omega_{DE}}\bigg) \\ \nonumber
+2 \bigg( \frac{\Omega_{DE}}{\Omega_m u+\Omega_{DE}} \bigg)^{1/6} \bigg] \bigg[ \ln \bigg( \frac{u}{\Omega_m u+\Omega_{DE}} \bigg)-2 \bigg] \\ \nonumber
-3 {_3 F_2} \bigg( \frac{1}{3}, \frac{1}{3}, \frac{5}{6}; \frac{4}{3}, \frac{4}{3}; \frac{\Omega_m u}{\Omega_m u+\Omega_{DE}} \bigg) \bigg\}  \bigg] \bigg/ \\ \nonumber
\left[\left(\frac{\Omega_{DE}}{\Omega_{m}u+\Omega_{DE}}\right)^{7/6}\left(\Omega_{m}u+\Omega_{DE}\right)^{3/2}\right].
\label{e:deltdcchi}
\end{eqnarray}

The second term on the right-hand side of eqn. (\ref{e:deltdcchi1}) comes from evaluating $I[u]$, which is the indefinite
integral resulting from eqn. (44), at the lower bound $z=0$. $_{2}F_{1}$ and $_{3}F_{2}$ are generalized hypergeometric functions.

Finally, we obtain
an expression for $d_{c,-1}$, the comoving distance in a cosmological constant cosmology.

\begin{equation}
d_{c,-1}\approx \frac{1}{3H_0} B_{-\frac{\Omega_{m}(1+z)^{3}}{\Omega_{DE}}}\left(\frac{1}{3},\frac{1}{2}\right)\left(-\Omega_{m}\right)^{-1/3}\Omega_{DE}^{-1/6}-\kappa,
\label{e:dcchi}
\end{equation}

with \[
\kappa\equiv B_{-\frac{\Omega_{m}}{\Omega_{DE}}}\left(\frac{1}{3},\frac{1}{2}\right)\left(-\Omega_{m}\right)^{-1/3}\Omega_{DE}^{-1/6}\]

and $B_{x}(a,b)$ the incomplete Beta function, given by\footnote{See Eisenstein 2013 for more pedagogical discussion of this calculation.}

\[
B_{x}(a,b)\equiv\frac{x^{a}}{a}\;_{2}F_{1}\left(a,1-b;a+1;x\right).\]

Combining eqns. (\ref{e:deltdcchi1})  and (\ref{e:dcchi}), we find

\begin{equation}
\frac{\Delta d_{c}}{d_{c,-1}}\approx\frac{\delta w_{0}}{2}\frac{I\left[(1+z)^{3}\right]-13}{B_{-\frac{\Omega_{m}(1+z)^{3}}{\Omega_{DE}}}\left(\frac{1}{3},\frac{1}{2}\right)\left(-\Omega_{m}\right)^{-1/3}\Omega_{DE}^{-1/6}-\kappa}.\end{equation}

The complexity of the expression is not particularly important; what is important is that is shows that $\Delta d_c/d_{c,-1}$ has a  very simple, linear dependence on $\delta w_0$, just as we earlier found for $\Delta H/H_{-1}$.  So we conclude

\begin{equation}
\frac{\Delta H}{H_{-1}}\propto\delta w_{0}\;\;\;{\rm and\;\;\;}\frac{\Delta d_{c}}{d_{c,-1}}\propto\delta w_{0}.\end{equation}

Now, the $\chi^{2}$ is just, with $DOF$ degrees of freedom,

 \begin{equation}
\chi^{2}=DOF+\sum_{z_{i}}\left(\frac{\Delta H/H_{-1}}{\sigma_{H}}\right)^{2}+\sum_{z_{i}}\left(\frac{\Delta d_{c}/d_{c,-1}}{\sigma_{d_{c}}}\right)^{2},
\label{e:chisqgeneralform}
\end{equation}

where $\sigma_H$ represents an observational error bar in $H$, and analogously for $\sigma_{d_c}$.  The sums over $z_i$ are simply sums over observations conducted at different redshifts.

Thus, it is evident that
 \begin{equation}
\chi^{2}-DOF\propto\delta w_{0}^{2}.
\label{e:chi2proptdeltaw}
\end{equation}

Now, we seek an analogous relation for $\chi^2$'s variation with changes in $\Omega_m$.  First, Taylor-expand $H_{-1}$ about $\Omega_m=.272$, since, the way we have set up our calculations (see \S5), we will be varying the value of $\Omega_m$ used to compute $H_{-1}$.
We use subscript $-1, \Delta \Omega_m$ to represent that $H$ here is calculated with $w \equiv -1$ DE but with a value of $\Omega_m$ different from the fiducial value of $.272$.  We have

\begin{equation}
H_{-1,\Delta\Omega_{m}}\approx H_{-1}+\frac{H_{0}^{2}}{2H_{-1}}\left[\left(1+z\right)^{3}-1\right]\Delta\Omega_{m},\end{equation}
 meaning

\begin{equation}
\frac{\Delta H}{H_{-1}}=\frac{H_{SR}}{H_{-1}}-1-\frac{H_{0}^{2}}{2H_{-1}^{2}}\left[\left(1+z\right)^{3}-1\right]\Delta\Omega_{m}.\end{equation}

Using the earlier Taylor series for $H_{SR}$ about $\delta w_0=0$ (eqn. (\ref{eq:Taylordeltaw})) in the first term on the right-hand side above and simplifying yields 

\begin{eqnarray}
\frac{\Delta H}{H_{-1}}= \frac{1}{2} \bigg( \frac{H_{0}}{H_{-1}} \bigg)^{2}\bigg\{  \ln\bigg[\frac{\big(1+z\big)^{3}}{\Omega_{m}\big(1+z\big)^{3}+\Omega_{DE}} \bigg]\delta w_{0} \\ \nonumber
-\big[\big(1+z\big)^{3}-1\big]\Delta\Omega_{m} \bigg\}.
\end{eqnarray}

We wish to extract a scaling for $\chi_H^{2}-DOF\propto\big( \Delta H_{\Delta\Omega_{m}}\big)^{2}$,
where we have written $\chi_H^{2}$ to denote that this is only true
for the contribution to $\chi^{2}$ that is from measurements of the
Hubble constant (see \S5.3 for further discussion). Making a number of approximations, we find that

\begin{equation}
\left(\Delta H_{\Delta\Omega_{m}} \right)^{2}\propto\left(1+\frac{\Delta\Omega_{m}}{A(u)\delta w_{0}}\right)^{2},
\label{e:deltHdeltm}
\end{equation}

where $A(u)\equiv\frac{1}{u^{3}-1}\ln\left[\frac{u^{3}}{\Omega_{m}u^{3}+\Omega_{DE}}\right]$
and $u\equiv\left(1+z\right)$. So this allows us to compute the change in $\chi^{2}$
from $H$ due to a change in the matter density, but only at
one particular fixed $z$ and $\delta w_{0}.$ In contrast, we would like to
calculate this change over all $z.$ Thus we evaluate $A$ at the average value of $u$,  $<u>=1+<z>$, where we take $<z>=1$ as this is the rough average redshift at which the observations we use are done. We therefore define $\alpha(\delta w_0)=1/\left(A(2)\delta w_{0}\right)$
and conclude that

\begin{equation}
\chi_H^2-DOF \propto \left[1+\alpha \left(\Delta \Omega_m \right) \right]^2,
\label{e:chisqscalingwithmatterappendix}
\end{equation}

where we have evaluated eqn. (\ref{e:deltHdeltm}) at $u=<u>$, used the definition of $\alpha$, and substituted the result into eqn. (\ref{e:chisqgeneralform}).

Finally, we consider how $\chi^2$ scales with changes in $H_0$.  Using methods similar to those detailed above, we easily find that 

\begin{equation}
\chi^2-DOF\propto (\Delta H/H_{-1})^2 \propto \bigg(- \bigg(\frac{H_{-1}}{H_0}\bigg) \Delta H_0+\gamma(u,\delta w_0) \bigg)^2,
\label{e:chisqvsh}
\end{equation}
where we have defined $\gamma(u, \delta w_0) = \frac {H_0^2 }{2 H_{-1}} \ln \big[\frac{u^3}{\Omega_m u^3+\Omega_{DE}} \big]\delta w_0$. In the limit that $\delta w_0$ is zero, $\chi^2$ is simply quadratic in $\Delta H_0/H_0$, as expected; when $\delta w_0\neq 0$, we may evaluate $\gamma$ at $<u>$ to obtain the appropriate scaling. Note that we may then also approximate that $H_{-1}$ in the denominator of the second term has the same value today as $H_{SR}$, corresponding to dropping a term in $\Delta H_0 \delta w_0/H_0$. We see that for non-zero $\gamma$, the scaling is not symmetric in $\Delta H_0/H_0$, so the sign of $\Delta H_0$ may matter as well as the magnitude.  Note that this scaling does not take into account changes in the matter density; we assume this is held fixed while changing $H_0$.  

Finally, we provide Figure 16 to show the effects of changes in the precision to which $\Omega_m$ is measured on the confidence of a detection of slow-roll DE.
\begin{figure}       
\includegraphics[scale=.7]{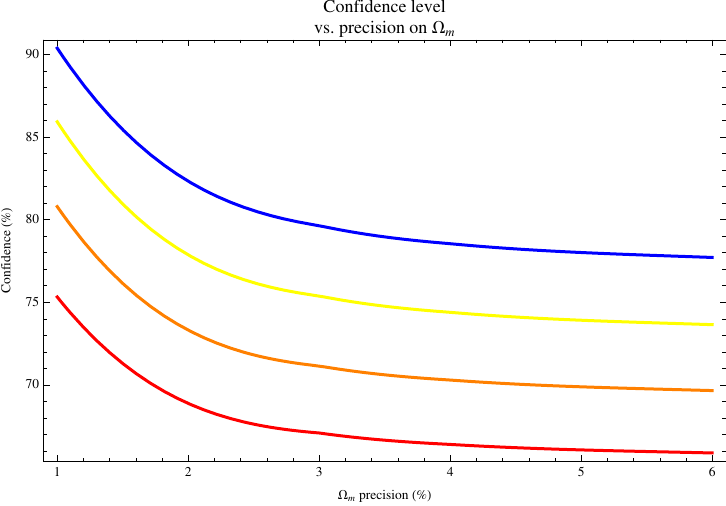} 
\centering{}\caption{These curves show how the confidence of a possible detection (assuming $1\%$ precision on $H_0$) changes as a function of the precision on $\Omega_m$.  From top to bottom the curves correspond to $\delta w_0=5\%,\;4.5\%,\;4\%$, and $3.5\%$.}
 \vspace{0.5in}
\label{f:confvsprecomml}
\end{figure}

\subsection{Reconstruction of confidence ellipse}
The general equation of an ellipse is
\begin{equation}
Ax^2+Bxy+Cy^2+Dx+Ey+F=0.
\label{e:ellipsestdform}
\end{equation}
For our analysis of Figure 8, we use three measurements from the confidence contour: the two axes, and the tilt angle $\theta$ between the contour's semi-major axis and the $\delta w_0$ axis.  So we will be unable to determine all six parameters above without some assumptions.  First, we take $D=0=E$; if we can derive the equation of an ellipse that fits the plot with this assumption, then that is all we require.  Second, we let $F$ be free; different values will correspond to different concentric ellipses in Figure 8, so $F$ can be considered an overall scaling we need not determine.  We thus have, writing eqn. (\ref{e:ellipsestdform}) with $D=E=0$ in matrix notation,
\[ 
\begin{array}{cc}
\left(x\;y\right)
\end{array}\cdot
\left( \begin{array}{cc}
A & B/2  \\
B/2 & A  
\end{array} \right)
\cdot
\left(x\atop y \right)=-F.
\] 

Diagonalizing the middle matrix, which we denote $\bf{Q}$, yields the decomposition

\begin{equation}
\bf{Q}=\bf{S}\bf{D}\bf{S}^{-1},
\end{equation}

with 

\[
\bf{S}=\left( \begin{array}{cc}
\frac{A-C-\sqrt{\lambda}}{B} & \frac{A-C+\sqrt{\lambda}}{B}  \\
1 & 1  
\end{array} \right)
\]

and

\[
\bf{D}=\left( \begin{array}{cc}
\frac{1}{2} (A+C-\sqrt{\lambda}) & 0  \\
0 & \frac{1}{2}(A+C+\sqrt{\lambda}),
\end{array} \right)
\]
with $\lambda \equiv (A-C)^2+B^2$.  $D_{11}$ corresponds to $1/a^2$, $a$ the semi-major axis of the ellipse, and $D_{22}$ corresponds to $1/c^2$, $c$ the semi-minor axis of the ellipse.  The column vectors that are the columns of $\bf{S}$ are the two principle axes of the ellipse, the longer one being represented by the second column of $\bf{S}$ as should be clear because $\lambda$ is positive. The angle formed by this vector with the $\delta w_0$ axis will be $\theta$, and so $\tan(\theta)=S_{22}/S_{12}$.  The system of equations given by $D_{11}=1/a^2$, $D_{22}=1/c^2$, and $\tan(\theta)=S_{22}/S_{12}$ may be solved to find that the ellipses in Figure 8 are given by
\[
.428\delta w_0^2-1.052\delta w_0\left(\Delta \Omega_m /\Omega_m\right)+1.319 \left(\Delta \Omega_m/\Omega_m\right)^2=-F
\]
where $-F>0$.

\subsection{Numerical method for \S10.2}
Here we briefly describe the method used to self-consistently numerically solve the Friedmann equation and the field equation of motion for the results presented in \S10.2.  We write the Friedmann equation in terms of $\tau \equiv tH_0$, and to avoid a singularity at $a=0$ multiply both sides by $a^2$.  This yields
\begin{equation}
 a \left( \frac{da}{d\tau} \right) =\left(\Omega_{m}a+\Omega_{r}+\Omega_{DE}[\phi,\phi']\right)^{1/2},
 \label{e:77}
\end{equation}
where here for clarity we have introduced subscripts ``nought'' to denote that the matter and radiation densities given are constants and evaluated today, at $a=1$.  Here we desire an exact numerical solution, so we account for the radiation energy density $\Omega_r$ though we neglect it in the rest of this work.  Thus the agreement of our exact numerical results with the formulae of \S2 also illustrates that this approximation is justified. The DE density is not a constant but is rather some function of $\phi$ and $\phi'\equiv d\phi/d\tau$, which themselves will be functions of $\tau$. The functional form changes depending upon the model being solved (the energy density for a quintessence field is different from that for a phantom field, for instance).

With the change of variable noted for time, the quintessence equation of motion is
\begin{equation}
\phi ''+3H(\tau)\phi'=-\frac{1}{H_0^2} \frac{\partial V}{\partial \phi};
\label{e:rescaledeom}
\end{equation}
the phantom field equation of motion is the same up to the sign in front of the right-hand side (see \S2.5 and eqn. (\ref{e:phantomeom})).  We have defined $H(\tau) = \frac{1}{a}\frac{da}{d\tau}$.  

We now have two coupled equations to solve (eqns. (\ref{e:77}) and (\ref{e:rescaledeom})); this is done using Mathematica's built-in NDSolve routine.  We set the initial conditions at a very small value of time and scale factor obtained from explicit numerical integration of the Friedmann equation with a cosmological constant DE; this is to avoid the singularity in $H(\tau)$ at $\tau=0$ otherwise encountered in eqn. (\ref{e:rescaledeom}).  So doing will introduce negligible error because DE is exceedingly subdominant to matter and radiation at the value we choose, $a=10^{-10}$.  

We specify initial conditions at $a=10^{-10}$ using the insight that, since the field should be in slow-roll, it will not change much from then until now. Thus, to obtain the correct DE density and desired $\delta w\simeq 5\%$ now, we may set the corresponding values of $V_0$ and $\phi'_0$ (at present)) as initial conditions at $a=10^{-10}$.  This first guess for the initial conditions does not yield precisely the correct DE density or $\delta w$ today, so we then change the initial conditions slightly and iterate until we achieve $\delta w\simeq 5\%$ and $\Omega_{DE}=\Omega_{DE0}$.  This process converges quickly and yields self-consistent, exact solutions to the coupled system.  These results are plotted in Figures 12-15. See also Gott \& Slepian 2011.

\clearpage
\section{Appendix B: Tables of Precisions Used}
\label{subsec:precisiontables}

Note that all precisions listed here are fractional precisions and not percentages.

\FloatBarrier

\begin{table}
 \vspace{0.2in}
\caption{WFIRST SNe Luminosity distance (optimistic) (Green et al. 2012).}
 \vspace{0.04in}
\centering
    \begin{tabular}{| l | c | c | c |}
 
    \hline
     z&Precision  \\ \hline
     0.17&0.018\\ \hline
     0.25&0.011\\ \hline
     0.35&0.009\\ \hline
     0.45&0.008\\ \hline
     0.55&0.008\\ \hline
     0.65&0.008\\ \hline
     0.75&0.008\\ \hline
     0.85&0.012\\ \hline
     0.95&0.012\\ \hline
     1.05&0.012\\ \hline
     1.15&0.012\\ \hline
     
     \end{tabular}
     \vspace{0.2in}
     \label{t:WFIRSTSNe}
\end{table}

\begin{table}
 \vspace{0.2in}
\caption{WFIRST Hubble constant (Green et al. 2012).}
 \vspace{0.04in}
\centering
    \begin{tabular}{| l | c | c | c |}
 
    \hline
     z&Precision  \\ \hline
     0.8&0.017\\ \hline
     0.9&0.014\\ \hline
     0.95&0.013\\ \hline
     1.1&0.012\\ \hline
     1.2&0.012\\ \hline
     1.3&0.012\\ \hline
     1.4&0.012\\ \hline
     1.45&0.012\\ \hline
     1.6&0.012\\ \hline
     1.7&0.012\\ \hline
     1.8&0.013\\ \hline
     1.9&0.014\\ \hline
     1.95&0.017\\ \hline
     
     \end{tabular}
     \vspace{0.2in}
     \label{t:WFIRSTH}
\end{table}

\begin{table}
 \vspace{0.2in}
\caption{BigBOSS Hubble constant (BigBOSS website).}
 \vspace{0.04in}
\centering
    \begin{tabular}{| l | c | c | c |}
 
    \hline
     z&Precision  \\ \hline
     0.15&0.039\\ \hline
     0.2&0.027\\ \hline
     0.4&0.021\\ \hline
     0.5&0.016\\ \hline
     0.6&0.014\\ \hline
     0.7&0.012\\ \hline
     0.75&0.01\\ \hline
     0.85&0.009\\ \hline
     1&0.009\\ \hline
     1.1&0.009\\ \hline
     1.2&0.012\\ \hline
     1.3&0.016\\ \hline
     1.4&0.017\\ \hline
     1.5&0.016\\ \hline
     1.6&0.017\\ \hline
     1.65&0.022\\ \hline

     \end{tabular}
     \vspace{0.2in}
     \label{t:BigBOSSH}
\end{table}

\begin{table}
 \vspace{0.2in}
\caption{Euclid Hubble constant (Refregier et al. 2010).}
 \vspace{0.04in}
\centering
    \begin{tabular}{| l | c | c | c |}
 
    \hline
     z&Precision  \\ \hline
     0.7&0.016\\ \hline
     0.8&0.016\\ \hline
     0.9&0.016\\ \hline
     1&0.016\\ \hline
     1.1&0.016\\ \hline
     1.2&0.016\\ \hline
     1.3&0.016\\ \hline
     1.4&0.017\\ \hline
     1.5&0.018\\ \hline
     1.6&0.021\\ \hline
     1.7&0.026\\ \hline
     1.8&0.032\\ \hline

     \end{tabular}
     \vspace{0.2in}
     \label{t:EuclidH}
\end{table}

\begin{table}
 \vspace{0.2in}
\caption{LSST BAO and WL comoving distance (Ivezic et al. 2011).}
 \vspace{0.04in}
\centering
    \begin{tabular}{| l | c | c | c |}
 
    \hline
     z&Precision  \\ \hline
     0.5&0.005\\ \hline
     0.8&0.005\\ \hline
     1.1&0.005\\ \hline
     1.4&0.005\\ \hline
     1.7&0.005\\ \hline
     2&0.005\\ \hline
     2.3&0.005\\ \hline
     2.6&0.005\\ \hline
     2.9&0.005\\ \hline

     \end{tabular}
     \vspace{0.2in}
     \label{t:LSSTBAOandWL}
\end{table}

\begin{table}
 \vspace{0.2in}
\caption{BOSS angular diameter distance (Eisenstein et al. 2011).}
 \vspace{0.04in}
\centering
    \begin{tabular}{| l | c | c | c |}
 
    \hline
     z&Precision  \\ \hline
     0.35&0.01\\ \hline
     0.6&0.011\\ \hline
     2.5&0.015\\ \hline
 
     \end{tabular}
     \vspace{0.2in}
     \label{t:BOSSdA}
\end{table}

\begin{table}
 \vspace{0.2in}
\caption{BOSS Hubble constant (Eisenstein et al. 2011).}
 \vspace{0.04in}
\centering
    \begin{tabular}{| l | c | c | c |}
 
    \hline
     z&Precision  \\ \hline
     0.35&0.018\\ \hline
     0.6&0.017\\ \hline
     2.5&0.015\\ \hline
 
     \end{tabular}
     \vspace{0.2in}
     \label{t:BOSSH}
\end{table}

\begin{table}
 \vspace{0.2in}
\caption{Precisions used on $\Omega_m$ and $H_0$.}
 \vspace{0.04in}
\centering
    \begin{tabular}{| l | c | c | c |}
 
    \hline
     Parameter&Precision  \\ \hline
      $\Omega_m$&0.0125\\ \hline
     $H_0$&0.01\\ \hline
 
     \end{tabular}
     \vspace{0.2in}
     \label{t:ommH0precis}
\end{table}


\end{document}